\documentclass[prc,showpacs,showkeys,amssymb,amsmath,twocolumn]{revtex4}
\usepackage{amssymb}
\usepackage{amsmath}
\usepackage[pdftex]{graphicx}
\usepackage{epsfig}
\usepackage{epstopdf}
\usepackage{color}
\usepackage{multirow}
\usepackage{slashed}
\newcommand{\be}{\begin{equation}}
\newcommand{\ee}{\end{equation}}
\newcommand{\ben}{\begin{eqnarray}}
\newcommand{\een}{\end{eqnarray}}

\begin{document}
\title{Properties of neutral mesons in a hot and magnetized quark matter: size-dependent effects}

\author{Luciano M. Abreu}
\email[]{luciano.abreu@ufba.br}
\affiliation{Instituto de F\'{\i}sica, Universidade Federal da Bahia, 40170-115, Salvador, BA, Brazil}

\author{Emerson B. S. Corr\^ea}
\email[]{emersoncbpf@gmail.com}
\affiliation{Faculdade de F\'isica, Universidade Federal do Sul e Sudeste do Par\'a,  68500-000, Marab\'a, PA, Brazil}

\author{Elenilson S. Nery}
\email[]{elenilsonnery@hotmail.com}
\affiliation{Instituto de F\'{\i}sica, Universidade Federal da Bahia, 40170-115, Salvador, BA, Brazil}
%
\begin{abstract}

In this work we investigate the combined finite-size and thermomagnetic effects on the properties of neutral mesons in a hot medium, in the context of the Nambu--Jona-Lasinio model. In particular, by using the mean-field approximation and the Schwinger proper time method in a toroidal topology with periodic and antiperiodic conditions, we evaluate the chiral phase transition, the critical coupling, the constituent quark mass and meson observables like the $\pi ^0$ and $\sigma$ meson masses and pion decay constant under the change of the size, temperature and strength of external magnetic field. The results indicate that the observables are strongly affected by the conjoint effects of relevant variables and also by the periodicity of the boundary conditions chosen, and the net result will depend on the balance of these competing phenomena.

\end{abstract}

\maketitle
%
\section{Introduction}

The comprehension of strongly interacting matter under extreme conditions remains nowadays as one of the major challenges for the experimental and theoretical particle physics community.  Theoretical studies have predicted its rich phase structure, with the emergence of a   deconfined state called quark-gluon plasma (QGP), already observed in heavy-ion collisions~\cite{rev-qgp,Prino:2016cni,Pasechnik:2016wkt}. However, a complete characterization of the phase diagram in the regime of intermediate temperature and chemical potential directly from its underlying theory, the Quantum Chromodynamics (QCD), is a very hard task. In light of the intrincate mathematical structure
of QCD, effective models that incorporate
some of its properties have been largely employed. 

In this sense,
four-fermion models, as the Nambu--Jona-Lasinio (NJL)
model, are very useful for the investigation of dynamical
chiral symmetry phase transition when the system is under extreme conditions~\cite{NJL,NJL1,Vogl,Klevansky,Hatsuda,Buballa}. 
It should be noticed that the properties of the chiral phase transition can be evaluated from the observables of the hadrons, especially those of the mesons, when they are in a hot and dense medium. Mesons are produced abundantly in collisions, are more susceptible to the medium conditions and have a direct way of association to the order parameters of the phase transition.

Another thermodynamic variable relevant for the assessment of the chiral phase diagram is the  magnetic field. In heavy-ion collisions and compact stars, a strong magnetic background is produced~\cite{Kharzeev,Skokov:2009qp,Chernodub:2010qx,Ayala1,Tobias,Heber,MAO,Ayala2,Mamo:2015dea,Pagura,Magdy,Zhang:2016qrl,Ayala0,Wang:2017vtn,Mao:2018dqe,Avancini:2018svs,Ghosh:2021dlo}. At RHIC and LHC, for example, a sizable dependence of the phase transition with the field strength $\omega = e H$ is estimated in the hadronic scale, i.e. $\omega \sim 1-15\; m_{\pi}^2$ ($ m_{\pi} = 135-140$ MeV is the pion mass). In this environment, interesting phenomena have been proposed by the use of effective approaches. For instance: magnetic catalysis (stimulation of broken phase) happens at smaller temperatures at sufficiently small $\omega$, while the inverse magnetic catalysis (restoration of chiral symmetry) appears at higher temperatures or even at $T=0$ for intermediate/high values of field strength~\cite{Tobias,MAO,Mamo:2015dea,Pagura,Magdy,Ayala0,Avancini:2018svs}.

On the other hand, finite-volume effects play also an important role  on the phase structure of strongly interacting matter. Estimations suggest that QGP-like systems produced in heavy-ion collisions have a finite volume of units or dozens of fm${^3}$, depending on the conditions (e.g. nuclei, energy and centrality)~\cite{Bass:1998qm,Palhares:2009tf,Graef:2012sh,Shi:2018swj}. Therefore, the size dependence of the thermodynamic behavior of quark gas and meson observables have been discussed largely via distinct effective approaches of QCD, including the NJL model~\cite{Shi:2018swj,Luecker:2009bs,Li:2017zny,Braun:2004yk,Braun:2005fj,Ferrer:1999gs,Abreu:2006,Ebert0,Abreu:2009zz,Abreu:2011rj,Bhattacharyya:2012rp,Bhattacharyya:2014uxa,Bhattacharyya2,Pan:2016ecs,Gasser:1986vb,Damgaard:2008zs,Fraga,Abreu3,Abreu6,Ebert3,Abreu4,Magdy:2015eda,Abreu5,Abreu7,Bao1,PhysRevC.96.055204,Samanta,Wu,Klein:2017shl,Shi,Wang:2018kgj,Abreu:2019czp,XiaYongHui:2019gci,Abreu:2019tnf,Das:2019crc,Zhao:2018nqa}.
The main conclusion is that thermodynamic properties of strongly interacting matter show dependence on finite-size effects. 
For the case of chiral symmetry phase transition, in the bulk approximation the system experiences a transition from chiral symmetry broken phase to the symmetric phase as the temperature and/or baryon chemical potential increases, with the quark-antiquark scalar condensate being interpreted as the order parameter. When the system is restricted to a finite volume, the chiral symmetric phase is stimulated as well. Thus, one may wonder about the conditions that an ideal bulk system seems a good approximation for systems constrained to boundaries. 

As a consequence of these last paragraphs, a natural discussion emerges: how are the phase structure of a hot quark gas and the meson properties affected when the system is simultaneously subject to boundaries and to a magnetic background. 
Hence, in the present work we intend to contribute to this debate and to address these questions. In particular, we will investigate the conjoint finite-size and thermomagnetic effects on the properties of neutral mesons in a hot medium, in the context of the two-flavor Nambu--Jona-Lasinio model. By using the mean-field approximation and the Schwinger proper time method in a toroidal topology with periodic (PBC) and antiperiodic (APBC)  boundary conditions, we analyze the gap equation solutions and meson observables like the $\pi ^0$ and $\sigma$ meson masses and pion decay constant under the change of the size, temperature and strength of external magnetic field. The finite size effects are implemented according to the generalized Matsubara prescription.

We organize the paper as follows. In Section~II, we calculate the $(T,L,\mu,\omega)$-dependent gap equation, the $\pi ^0$ and $\sigma$ meson masses and pion decay constant from the NJL model in the mean-field approximation, using Schwinger's proper-time method and generalized Matsubara prescription. The results concerning the phase structure of the system, the critical coupling, the behavior of constituent quark mass, the spatial and chiral susceptibilities and the meson properties are shown and analyzed in Section~III. Finally, Section~IV presents some concluding remarks.


\section{The NJL model with boundaries and a magnetic background}

\subsection{NJL model and meson properties}
	
We begin presenting the NJL model, whose Lagrangian density is given by~\cite{NJL,NJL1,Vogl,Klevansky,Hatsuda,Buballa}
\begin{eqnarray}
\mathcal{L}_{ NJL} & = & \bar{q}\,(i{\slashed{\partial}} - \hat{m} )\,q +  g_{s} \, \left[\left(\bar{q}q\right)^{2}+\left(\bar{q}\,i\gamma_{5}\vec{\tau}\,q\right)^{2}\right] 
,
\label{L}
\end{eqnarray}
where $q = (u,d)^{T}$ denotes the light quark field doublet $ (N_f = 2) $ with $ N_c = 3 $ color, and $ \bar{q} $ its respective antiquark field doublet; $\hat{m} = {\rm diag}(m_{u},m_{d})$ is the current quark mass matrix; $g_{s} $ is the coupling constant of the scalar and pseudoscalar channel, and $\vec{\tau} $ are the generators of $U(2)$ in flavor space (i.e. the Pauli matrices). 
We assume henceforth the isospin symmetry on the Lagrangian level, i.e. $ m_u = m_d \equiv m $, and therefore $\hat{m} = m \,\mathbf{1}$ . 

The present study is devoted to the lowest-order estimate of the phase structure and meson properties, and the calculations are performed in the context of the mean-field (Hartree) approximation. Accordingly, the quark condensate is the only allowed expectation value bilinear in the quark fields: $ \phi \equiv \left\langle  \bar{q} q \right\rangle  $. 
Therefore, the interaction terms in $\mathcal{L}_{ NJL}$ are linearized in the presence of $\phi $: $  (\bar{q} q) ^2 \simeq 2 \phi  (\bar{q} q) - \phi  ^2 $; other terms quadratic in the fluctuations will be neglected, and those in channels without condensate or nondiagonal in flavor space are excluded. 
This allows us to obtain the following gap equation,
\be
M = m + 2 N_f {g_s} \phi,
\label{massa}
\ee 
where we have introduced $M $ as the constituent quark mass. 
The physical solutions of Eq.~(\ref{massa}) are determined from the stationary points of the effective potential, which lead to the standard expression for the quark condensate,
\be
\phi \equiv \left\langle  \bar{q}_f q_f \right\rangle = \mathrm{i}\int \frac{d^4 p}{(2 \pi)^4} \mathrm{Tr}(S(p)), 
\label{phi1-ini}
\ee
where $\mathrm{Tr}$ means the trace over Dirac and color spaces, and $ S(p) $ the $q_f$-quark propagator, 
\be
S(p) = \frac{\slashed{p}+M}{({p}^2-M^2 )}.
\label{propagator1}
\ee
The chiral condensate in Eq.~(\ref{phi1-ini}) can be rewritten in a more convenient form,
\be
\phi =  4\,\mathrm{i}  N_c M I_{1},
\label{phi}
\ee
with 
\be
I_{1} = \int \frac{d^4 q}{(2 \pi)^4} \frac{1}{q^2 - M^2 }.
\label{I1}
\ee

In the context of the NJL model, the mesons can be interpreted as bound states of a dressed-quark and a dressed-antiquark, and analyzed via the Bethe-Salpeter equation (BSE) formalism by making use of the random phase approximation (RPA)~\cite{Klevansky}. In this approximation, the solution to the BSE for a given meson $ \alpha $ is given by a two-body correlation function defined as
\be
D_{\alpha} (p) = \frac{2g_{\alpha}}{1-2 g_{\alpha} \Pi_{\alpha}(p)}, 
\label{corrfunc} 
\ee
where $g_{\alpha}$ is the appropriate coupling constant and $ \Pi_{\alpha}(p) $ is the one-loop polarization function for the meson channel considered, written as 
\be
\Pi_{\alpha}(p) = -\mathrm{i} \int \frac{d^4 q}{(2\pi)^{4}} {\mathrm{Tr}}\left[iS(q+p)\Gamma_{\alpha}iS(q)\Gamma_{\alpha}\right],
\label{Pol1} 
\ee
with $\Gamma_{\alpha} = (I_{4},i\gamma_{5}\vec{\tau})$ denoting the adequate combination of gamma matrices of a specific meson channel of type $\alpha$. 

Focusing attention on the cases of $ \pi $ and  $ \sigma $ 
mesons, then $\Gamma_{\alpha}$ can be identified as $ I_{4},i\gamma_{5}\vec{\tau}  $, respectively. Using them in Eq. (\ref{Pol1}), after some manipulations the polarization functions are~\cite{Zhang:2016qrl}
\begin{eqnarray}
\Pi_{\sigma}(p) & = &  4\mathrm{i}N_c N_f\left[ I_{1} -\frac{1}{2}(p^2 - 4 M^2) I(p)\right], \nonumber \\ 
\Pi_{\pi}(p) &  = & 4\mathrm{i}N_c N_f \left[ I_{1}-\frac{1}{2}p^2  I(p)\right],
\label{Pol2}
\end{eqnarray}
where $I(p) $ means the scalar one-loop two-point function
\begin{eqnarray}
I(p) = \int \frac{d^4 q}{(2\pi)^4} \frac{1}{[(q+p)^2 - M^2][q^2 - M^2]}.
\label{Intp}
\end{eqnarray}

The mesons masses are identified from the pole mass position conditions in the respective correlation functions in Eq. (\ref{corrfunc}), namely:
\be
1-2 g_\alpha \Pi_{\alpha}(m_{\alpha}) = 0.
\label{GapEq.Mesons}
\ee

Besides, the pion decay constant $f_{\pi} $ can be calculated from the vacuum to one-pion axial vector matrix element, $ \langle 0 | \bar{q} \gamma ^{\mu} \gamma _5 \frac{\tau ^a}{2} q | \pi ^b (p) \rangle = i \delta ^{ab} f_{\pi} p_{\mu}$. Some manipulations of this element engender a suitable expression for $f_{\pi} $, 
\begin{eqnarray}
f_{\pi} ^2 = - 4 i N_c M^2 I(0), 
\label{piondecay} 
\end{eqnarray}
with $ I(p) $ defined in Eq. (\ref{Intp}). 

Noting that our interest is on the influence of combined effects of temperature, finite baryon density, magnetic magnetic and boundaries in the meson properties, it is convenient  to invoke the Schwinger proper-time approach~\cite{Schwinger,DeWitt1,DeWitt2,Ball} in order to treat the functions $I_{1}$ and $I(p)$ in a pragmatic manner. Firstly, we consider $I_{1}$ in Eq.~(\ref{I1}). After performing the Wick rotation, i.e. $q = (q_{t},\vec{q}\,) \rightarrow q_{E} = (\mathrm{i}q_{\tau},\vec{q}\,) $, and appplying the standard method of Schwinger, the integral $I_{1}$ reads
\begin{eqnarray}
I_1 & =&  -\mathrm{i}\int \frac{d^4 q_E}{(2\pi)^4} \frac{1}{q^2_{E} + M^2} 
 \nonumber \\
 & = &  - \mathrm{i}\int \frac{d^4 q_E}{(2\pi)^4} \int_{0}^{\infty}dS\exp\left[-(q_{E}^{2}+M^2)S\right],
\label{I1E}
\end{eqnarray}
where $S$ is the proper time and $q_{E}^{2} = q_{\tau}^{2}+q_{x}^{2}+q_{y}^{2}+q_{z}^{2}$.

In the case of the function $I(p)$ in Eq. (\ref{Intp}), starting with the use of the identity usually called as Feynman parametrization, 
\begin{eqnarray*}
\frac{1}{AB} = \int_{0}^{1}dx \int_{0}^{\infty}dS \,S\, e^{[(xA+(1-x)B)S]},
\end{eqnarray*}
it can be expressed as
\begin{eqnarray}
I(p) = \int \frac{d^4 q}{(2\pi)^4} \int_{0}^{1}dx \int_{0}^{\infty}dS\,S\,e^{[(x p^2 + 2  x \, q \cdot p +q^2 -M^2)S]}.
 \nonumber \\
\label{Intp1}
\end{eqnarray}
In the following, we define $p$ as a time-like vector and execute the Wick rotation in momenta space, namely $p^{\mu} \equiv (m_\alpha,\vec{0}\,) \rightarrow (\mathrm{i}m_\alpha,\vec{0}\,)$, yielding 
\begin{eqnarray}
I(m_\alpha) &=& \mathrm{i}\int \frac{d^4 q_E}{(2\pi)^4} \int_{0}^{\infty}dS\,S\,\int_{0}^{1}dx \,  e^{[-(x-x^2)m_{\alpha}^{2}S]} \nonumber \\
&\times& e^{\left\{-[(q_{\tau} + x \, m_{\alpha})^{2}+\vec{q}^{\,2} + M^2]S\right\}}.
\end{eqnarray}
Afterwards, we change the variable $(q_{\tau}+ x \, m_{\alpha})\rightarrow q_{\tau}$ and make the integral in $x$, resulting in the expression
\begin{eqnarray}
I(m_\alpha) &=& \mathrm{i}\int \frac{d^4 q_E}{(2\pi)^4} \int_{0}^{\infty}dS\,S\,  e^{\left[-\frac{m_{\alpha}^{2}S}{4}\right]}\left(\frac{\sqrt{\pi}}{m_{\alpha} \sqrt{S}}\right)
 \nonumber \\
 &  & \times Erfi\left(\frac{m_{\alpha} \sqrt{S}}{2}\right) 
 e^{\left[-(q^{2}_{E} + M^2)S\right]},
\label{ImE}
\end{eqnarray}
where $Erfi(a)$ is the imaginary error function. 

\subsection{Generalized Matsubara prescription}

Until now the discussion has been devoted to the situation of the system at zero temperature and baryon density, as well as in absence of spatial boundaries and a magnetic background. 
Now, we can apply the quantum field theory in a toroidal topology to include finite temperature, density and size effects. 

We take the Euclidean coordinates in Eqs.~(\ref{I1}) and (\ref{Intp}), denoted by $x_E = (x_{\tau},x_1,x_2,x_3)$, compactified as follows: $x_{\tau}\in[0,\beta]$ and $x_j\in[0,L_j] \; (j=1,2,3)$ , with $ \beta = 1/T$ being the inverse of temperature $T$ and $L_j$ the length of the compactified spatial dimensions. Consequently, the Feynman rules must be modified according to the so-called generalized Matsubara prescription~\cite{livro,Abreu:2003zz,Abreu:2003es,PR2014,Emerson},
\begin{eqnarray}
\int\frac{d^4 q_{E}}{(2\pi)^4}\,h(q_{\tau},\vec{q}\,)\rightarrow \frac{1}{\beta L_1 L_2 L_3}\sum_{ \{n_{i} \}=-\infty}^{\infty} h \left( \{\tilde{\omega}_{n_{i}} \} \right),\label{feynmanrule}
\end{eqnarray}
$(i= \tau, 1,2,3)$ such that
\begin{eqnarray}
  {q}_{\tau} & \rightarrow & \tilde{\omega}_{n_\tau} \equiv \frac{2\pi}{\beta}
	\left(n_{\tau}+\frac{1}{2} - i\frac{\mu \beta}{2\pi} \right),  \nonumber \\ 
 {q}_{j} & \rightarrow &  \tilde{\omega} _{n_j} \equiv \frac{2\pi}{L_{j}}
	\left(n_{j}-b_{j}\right) \, ,  \label{Matsubara}
\end{eqnarray}
where $n_{\tau}, n_{j} = 0,\pm 1 , \pm 2, \cdots$. 
Due to the fermionic nature of the system, the Kubo-Martin-Schwinger conditions~\cite{livro} require APBC in the imaginary-time coordinates. 

Concerning the spatial compactified coordinates, it is worthy mentioning that there are no conceptual restrictions regarding their periodicity, as stressed in several works~(\cite{Ferrer:1999gs,Isham:1977yc,Ishikawa:1996jb,Klein:2017shl,Abreu:2020uxc}). This choice depends on the physical interest. That being so, the parameters $b_{j}$ in Eq.~(\ref{Matsubara}) can assume the values 0 or $-1/2$ for PBC and APBC, respectively, and reverberate on the physical quantities obtained in the effective approach. We mention, for instance, the aspect with respect to the spacetime
permutation symmetry: for APBC in spatial compactified coordinates, the fermionic nature of the quark field engenders the physical equivalence of Euclidean space and time directions, keeping the permutation symmetry among them. As a consequence, taking the vacuum coupling constants of the model temperature-independent, then permutation symmetry assures that they do not depend on the size of spatial compactified coordinates. In the opposite way, the periodic condition PBC breaks this permutation symmetry, and therefore such a dependence cannot be eliminated a priori (we refer to Ref.~\cite{Klein:2017shl} for a detailed discussion). In next Section the physical meaning of the boundary conditions on the thermodynamic properties and the effective masses of the system will be examined.

To perform the manipulations of Matsubara series in a more tractable way, we employ the Jacobi theta functions $\theta_{2}(z;q)$ and  $\theta_{3}(z;q)$~\cite{Bellman,Mumford}, given by 
\begin{eqnarray}
\theta_{2}(u;v) & = & 2 \sum_{n=0}^{\infty} v^{(n+1/2)^2}\cos[(2n+1)u], \nonumber\\
\theta_{3}(u;v) & = & 1 + 2 \sum_{n=1}^{+\infty} v^{n^2}\cos(2nu).
\label{Theta} 
\end{eqnarray}
Therefore, the use of Eq.~(\ref{Matsubara}) and~(\ref{Theta}) allows to rewrite Eq.~(\ref{I1E}) as
\begin{eqnarray}
I_{1}(\beta,\mu,L_j) &=& -\mathrm{i}\frac{1}{\beta L_{1}L_{2}L_{3}}\int_{0}^{\infty} dS \exp[-S(M^{2}-\mu^{2})]
\nonumber \\
 &  & \times \,\theta_{2}\left[\frac{2\pi\mu S}{\beta}\,;\,\exp\left(-\frac{4\pi^2 S}{\beta^2}\right)\right] \nonumber \\
&&\times \prod_{j=1}^{3} \theta_{BC}\left[0\,;\,\exp\left(-\frac{4\pi^2 S}{L_{j}^2}\right)\right], 
\label{I1ETL}
\end{eqnarray}
where $  \theta_{BC} \equiv  \theta_{2} $ or $ \theta_{3}$ for APBC or PBC in spatial coordinates, respectively. 


Proceeding similarly with the function $I(m_{\alpha})$ in Eq.~(\ref{ImE}), we get
\begin{eqnarray}
I(m_\alpha,\beta,\mu,L_{j}) &=& \mathrm{i} \frac{1}{\beta L_{1}L_{2}L_{3}} \int_{0}^{\infty}dS\,S\, e^{\left[-\frac{m_{\alpha}^{2}S}{4}\right]}
\nonumber \\
 &  & \times \left(\frac{\sqrt{\pi}}{m_{\alpha} \sqrt{S}}\right)\,Erfi\left(\frac{m_{\alpha} \sqrt{S}}{2}\right) \nonumber \\
& & \times e^{[-S(M^{2}-\mu^{2})]}\,\theta_{2}\left[\frac{2\pi\mu S}{\beta}\,;\,e^{\left(-\frac{4\pi^2 S}{\beta^2}\right)}\right] 
\nonumber \\
 &  & \times \prod_{j=1}^{3} \theta_{BC}\left[0\,;\,e^{\left(-\frac{4\pi^2 S}{L_{j}^2}\right)}\right]. 
\label{ImETL}
\end{eqnarray}

%
%


\subsection{Inclusion of a magnetic background}

Here we present the model by considering the system under the influence of an external magnetic field. The magnetic background is implemented by means of minimal coupling prescription, where the ordinary derivative in Eq.~(\ref{L}) is replaced by the covariant one, that is $\partial_{\mu} \rightarrow \partial_{\mu} + \mathrm{i} \hat{Q} A_{\mu}$, where $A_{\mu}$ is the four-potential associated to the external magnetic field ${\bf{H}} = H \hat{z}$, and $\hat{Q}$ is the quark charge electric matrix, $ \hat{Q} = \mathrm{diag}\left( Q_u, Q_d \right) e $, with $Q_u = - 2 Q_d = 2 /3$. The choice of Landau gauge $A^{\mu } = (0,0,xH,0)$ yields a constant magnetic field $H$ along to $z$ direction.
Then, the gap equation in Eq.~(\ref{massa}) is reexpressed as  
\be
M = m + 2 {g_s} \sum_{f =u,d}  \phi_f (\omega),
\label{massaH}
\ee 
where the magnetic-dependent chiral condensate $\phi_f (\omega) $ is given by 
\be
\phi_f(\omega) =  \mathrm{i} \left(\frac{|Q_f| \omega}{2\pi}\right) \int \frac{d q_0 d q_3}{2 \pi} \sum_{\ell = 0}^{\infty} \sum_{s = \pm 1}^{}  \mathrm{Tr}(S_f(q,\omega)),
\label{condensate2}
\ee
with the quark propagator $S_f(q,\omega)$, connected to the inverse of the eigenvalues of the modified differential operator, taking the form
\be
S_f(q,\omega) = \frac{\slashed{q} - Q_f e \slashed{A}+M}{[{q}_{0}^{2}-q_{z}^{2}-|Q_f| \omega (2\ell+1-s)-M^2 ]},
\label{propagator2}
\ee
In equations above $\omega \equiv e H $ denotes the cyclotron frequency, $ s=\pm 1$ the spin polarization  and $\ell$ the Landau levels.

Now we can apply the recipe presented in the previous section. Making the Wick rotation and considering a Schwinger proper time parametrization, the expression for the chiral condensate is similar to the one in Eq.~(\ref{phi}), i.e. $\phi_f =  4\,\mathrm{i}  N_c M I_{1f}(\omega)$, but with $I_{1f}(\omega) $ being the magnetic-dependent function
\begin{eqnarray}
I_{1f}(\omega) & = & -\mathrm{i} \left(\frac{|Q_f| \omega}{2\pi}\right) \sum_{\ell = 0}^{\infty} \sum_{s = \pm 1}^{} \int \frac{d^2 q_{E}}{(2 \pi)^2} \int_{0}^{\infty} dS 
\nonumber \\
 &  &  \times  e^{[-(q_{E}^2+M^2)S]} \,e^{\left[- |Q_f |\omega \left(2 \ell + 1 - s \right) S\right]}.
\label{I1fprev}
\end{eqnarray}
Here $q_{E}$ is defined as $q_{E} = (q_{\tau}, q_{3})$. 
To include finite temperature, chemical potential  and size effects, we proceed analogously to the case without external field derived before and use Matsubara generalized prescription ~(\ref{Matsubara}). The resulting expression can be written, after performing the sum over the spin polarizations $s$ and the geometrical series in $\ell$, as 
\begin{eqnarray}
I_{1f}(\omega,\beta,\mu,L_3) &=& -\mathrm{i} \left(\frac{|Q_f| \omega}{2\pi \beta L_{3}}\right) \int_{0}^{\infty} dS e^{[-(M^2 - \mu^2)S]}
\nonumber \\
 &  & \times \theta_{2}\left[\frac{2\pi\mu S}{\beta}\,;\,e^{\left(-\frac{4\pi^2 S}{\beta^2}\right)}\right]
 \nonumber \\
 &  & \times \theta_{BC}\left[0\,;\,e^{\left(\frac{-4\pi^2 S}{L_{3}^2}\right)}\right] \, \coth{( |Q_f| \omega S)}. \nonumber \\
\label{I1fTL}
\end{eqnarray}

Now we take into account the combined thermomagnetic and finite-size effects on the meson properties. First we should mention that the polarization functions for pions 
include contribution of $u,d-\bar{u},\bar{d}$ quark-antiquark loops, and in absence of a magnetic background we have no distinction among the the polarization functions of neutral and charged mesons. In this sense, in most of the cases the choice of parameters is done using the neutral mesons. 
In the presence of an external magnetic field, the quark loop structure must be different for the charged mesons, and the  $\Pi_{\pi,\rho} (p)$  assumes a different form, while in the case of neutral mesons they are similar to Eq.~(\ref{Pol2}), keeping the modifications due to the dimensional reduction in the functions $I_1$  and $I(p)$. Thus, bearing in mind that our purpose here is a global analysis of the $\omega,\beta,\mu,L$-dependence of the relevant observables, here we focus on the neutral mesons. 

Therefore, we implement the $\omega,\beta,\mu,L$-dependence in the polarization functions for the $ \sigma, \pi^0$ mesons 
by performing in Eq.~(\ref{Pol2}) the following substitutions 
\begin{eqnarray}
(N_f I_1)  & \rightarrow  & \sum_{f =u,d}   I_{1f}(\omega,\beta,\mu,L_3), \nonumber \\
 (N_f I(p))  & \rightarrow &  \sum_{f =u,d}  I_{f}(p,\alpha,\omega,\beta,\mu,L_3),
\label{subs} 
\end{eqnarray}
where $ I_{1f}(\omega,\beta,\mu,L_z) $ is given in Eq.~(\ref{I1fTL}). The function $ I_{f}(\omega,\beta,\mu,L_z) $ is obtained from Eq.~(\ref{Intp}), first performing the Wick rotation, Feynman and Schwinger proper time parametrization, giving 
\begin{eqnarray}
I_{f}(m_\alpha,\omega) &=& \mathrm{i} \left(\frac{|Q_f| \omega}{2\pi}\right) \sum_{\ell = 0}^{\infty} \sum_{s = \pm 1}^{} \int \frac{d^2 q_{E}}{(2 \pi)^2}  \int_{0}^{\infty}dS\,S\, 
\nonumber \\
& & \times e^{\left[-\frac{m_{\alpha}^{2}S}{4}\right]} \left(\frac{\sqrt{\pi}}{m_{\alpha} \sqrt{S}}\right)\,Erfi\left(\frac{m_{\alpha} \sqrt{S}}{2}\right) \nonumber \\
& & \times e^{\left[-(q^{2}_{E} + M^2)S\right]} e^{\left[- |Q_f |\omega \left(2 \ell + 1 - s \right) S\right]}.
\label{ImEH}
\end{eqnarray}
After, with the use of the Matsubara prescription and the sum over the spin polarizations $s$ and the geometrical series in $\ell$, we obtain the final expression 
\begin{eqnarray}
I_{f}(m_\alpha,\omega,\beta,\mu,L_3) &=& \mathrm{i} \left(\frac{|Q_f| \omega}{2\sqrt{\pi} \beta L_{3}}\right)  \int_{0}^{\infty}dS\,\sqrt{S}\, e^{\left[-\frac{m_{\alpha}^{2}S}{4}\right]}\nonumber \\
& & \times \left(\frac{1}{m_{\alpha} }\right)\,Erfi\left(\frac{m_{\alpha} \sqrt{S}}{2}\right) \nonumber \\
& & \times e^{[-(M^2 - \mu^2)S]} 
\nonumber \\
& & \times 
\theta_{2}\left[\frac{2\pi\mu S}{\beta}\,;\,e^{\left(-\frac{\pi^2 S•}{\beta^2}\right)}\right]  \nonumber \\
& & \times\theta_{BC}\left[0\,;\,e^{\left(-\frac{4\pi^2 S}{L_{3}^2}\right)}\right] 
\nonumber \\
& & \times   \coth{(|Q_f| \omega S)} .
\label{ImEHL}
\end{eqnarray}

Finally, the thermal-magnetic-density-size dependence on the pion decay constant is determined from the expression 
\begin{eqnarray}
f_{\pi} ^2 = - 4 i N_c M^2  \sum_{f =u,d} I_{f}(0,\omega,\beta,\mu,L_z), 
\label{piondecay2} 
\end{eqnarray}
with $ I(p) $ defined in Eq. (\ref{ImEHL}).

A last remark is concerning the regularization procedure adopted. We deal with the divergences appearing at $S \rightarrow 0 $ by implementing an ultraviolet cutoff $\Lambda$ in the integral over $S$~\cite{Schwinger,Abreu:2019czp}
\begin{eqnarray}
\int_{0}^{\infty}\,h(S)\,dS \rightarrow \int_{1/\Lambda^2}^{\infty}\,h(S)\,dS.
\label{cuttof}	
\end{eqnarray}

\section{Results}

In this section we concentrate our attention on how the relevant quantities introduced above behave with the change of the  thermodynamic variables and, in particular, on the influence of the boundaries on the behavior of meson properties.  We simplify the present study by fixing $L_{i}=L$. Another important ingredient relies on our main intention of applying  our findings to the heavy-ion collision environment, which is characterized by a very low $\mu$. Therefore, we concentrate attention on the influence of thermodynamic variables $T, 1/L$ and $\omega$.

The NJL model with isospin symmetry in mean-field parameters is characterized by the following parameters: the coupling constant $ g_s $, the ultraviolet cutoff $ \Lambda$, and the current quark mass $ m$ or equivalently the constituent quark mass $M$. They are set in order to reproduce the observable hadron quantities at vacuum values of thermodynamic characteristics: $T,  1/L, \omega = 0$ (we refer the reader to Ref.~\cite{Klevansky}  for a detailed discussion; see also~\cite{Kohyama:2016fif} for the 3-flavor model). 
Actually, this limit is reached in our numerical calculations by taking the region of $(T,  1/L, \omega)$ in which they are sufficiently small to generate  the convergence to the vacuum values of the observables (see discussion in next Subsection). 
Specifically, the parameters are fixed by fitting the pion mass $m_{\pi} = 0.135~ \mathrm{GeV}$, and pion decay constant $f_{\pi} = 0.092~ \mathrm{GeV}$. So, the values we have obtained and used in the estimations are: $M = 0.300\,\mathrm{GeV} $,  $ g_s = 5.691~ \mathrm{GeV}^{-2}$, and $ \Lambda = 0.688~ \mathrm{GeV}$. This set engenders the current quark mass $m = 11.7~ \mathrm{MeV}$~\footnote{The value fixed for $m$ is higher than those usually utilized in different parametrizations (usually of the order of 5 MeV). This comes from the fact that most of authors prefer set the current quark mass close to the value estimated by quark models (the discussion on the quark masses in PDG's review). However,  there are also several works  which choose  the specification of the value of the constituent quark mass $M$, see for example~\cite{Cloet:2014rja,Zhang2016MPLA,Hutauruk:2018zfk}; besides, Ref.~\cite{Kohyama:2016fif} also presents other parametrizations in Tables 1-6 with higher values of $m$ in the context of 3-flavor model. The key role in this kind of effective model is that we have the freedom to specify the current quark mass or the constituent quark mass, since one determines the other through the gap equation; but in the end the set of parameters chosen must reproduce the physical quantities, i.e. the meson properties. 
}.

\subsection{The critical coupling}

We start by analyzing the critical coupling $g_s ^{(c)}  $ in parameter space that delimits the regions where the chiral symmetry is preserved or broken. On this issue we follow the discussion done in Ref.~\cite{Martinez:2018snm}. In the vacuum, it can determined from nontrivial solutions $M \neq 0$ of the gap equation (\ref{massa}) at chiral limit ($m=0$). In another perspective, it might be found from the intersections of the curve $f(y) = 2 N_f g_s \phi (y)$ and the line $y=M$. 
According to the magnitude of $g_s$, it could acquire a critical  value $g_s ^{(c)}  $ where exists one more intersection  beyond the trivial case $M=0$, and  a nontrivial solution $M\neq 0$ appears. In this sense, for $g_s > g_s ^{(c)}  $ at  least  two  intersections  are  found. Thus, the value of $g_s ^{(c)}  $ at which the  trivial and  nontrivial  solutions  bifurcate  from  one  another can be identified from the derivative of the  gap  equation with respect  to $M$ at $M=0$. This recipe for the vacuum scenario gives  $g_s ^{(c)} = 2 \pi^2 / (3 N_f \Lambda^2) $.

It is interesting to understand the behaviour of critical coupling by taking thermomagnetic and finite-size effects. In this case, the condition  for  criticality is obtained by applying the prescription described above in the bulk vacuum to the modified gap equation (\ref{massaH}), and reads
\ben
1 = 8 i N_c  {g_s ^{(c)} } \sum_{f =u,d} \left. \frac{\partial }{\partial M}  \left[ M I_{1f}(\omega,\beta,L)  \right] \right|_{M=0} ,
\label{critcond}
\een 
where $I_{1f}(\omega,\beta,L) $ is given in Eq.~(\ref{I1fTL}), and $g_s ^{(c)} $ now denotes  the  dressed critical  coupling  in  the  thermomagnetic  medium  with boundaries for light quarks to  have the constituent quark mass larger than the current quark mass~\cite{Martinez:2018snm}. 
Then, Eq.~(\ref{critcond}) can be interpreted as a self-consistent relation for the thermodynamic variables $(L, T, \omega) $ needed to break chiral symmetry. Its solution yields the value  $g_s ^{(c)} ( T, 1/L, \omega \rightarrow 0 ) \approx 3.46 \, \mathrm{GeV}^{-2}$  for the cutoff introduced at beginning of this Section. But keeping in mind that the heavy-ion collision ambience is characterized by high temperatures, then the solution of Eq.~(\ref{critcond}) for example for $T= 175 $ MeV gives $g_s ^{(c)} ( T = 175 \, \mathrm{MeV}; 1/L, \omega \rightarrow 0 ) \approx 5.93 \, \mathrm{GeV}^{-2}$. As a consequence, the coupling constant we chose  $(g_s = 5.691 \, \mathrm{GeV}^{-2})$ is most likely subcritical in the regime of high temperatures, where the chiral symmetry  is not broken, in accordance with our purposes of heavy-ion collision applications.

\begin{figure}
\centering
\includegraphics[width=1\columnwidth]{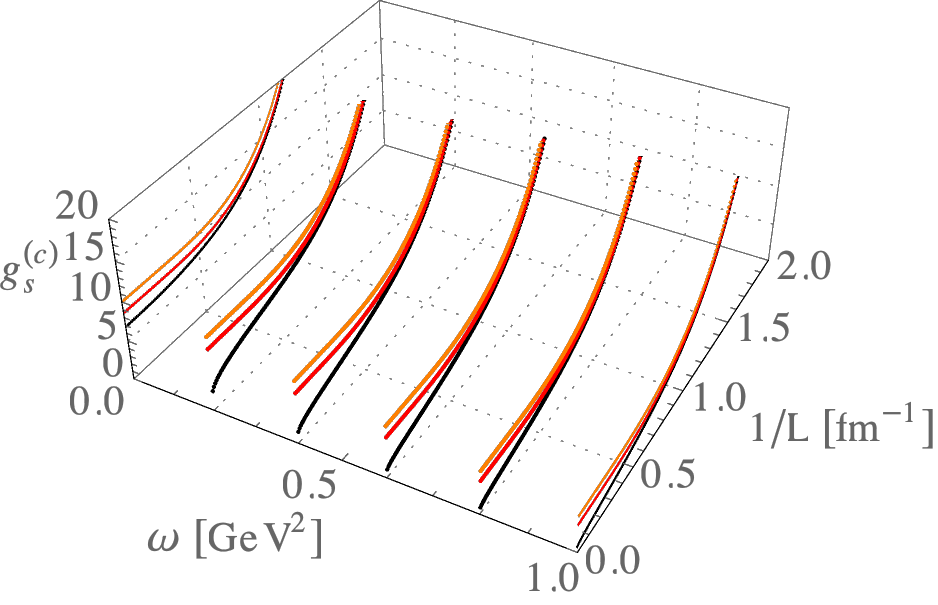}
\caption{Critical curves obtained from Eq.~(\ref{critcond}) as  a  function  of  the inverse of length $ 1/L $  in APBC case and cyclotron frequency $ \omega $, taking different values of temperature $T$. The critical coupling is given in units of GeV${}^{-2}$. The black, red and orange points represent the results for $T = 0, 0.120$ and $0.150$ GeV, respectively.}
\label{CriticalCouplingAPBC}
\end{figure}

\begin{figure}
\centering
\includegraphics[width=1\columnwidth]{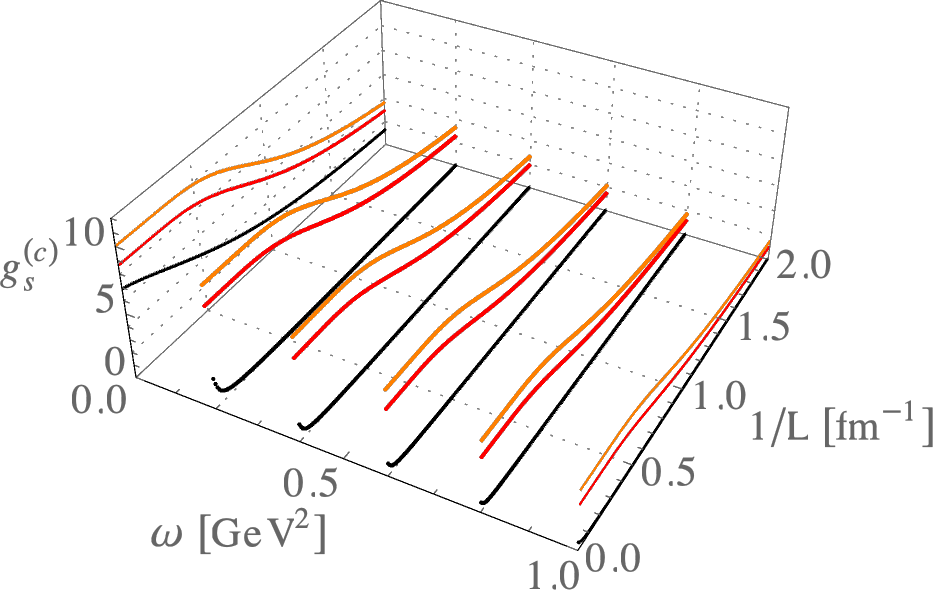}
\caption{The same as in Fig.~\ref{CriticalCouplingAPBC}, but in PBC case.}
\label{CriticalCouplingPBC}
\end{figure}

To have a more accurate comprehension of this point, in  Figs.~\ref{CriticalCouplingAPBC} and \ref{CriticalCouplingPBC} we  plot the critical curves obtained from Eq.~(\ref{critcond}) as  a  function  of  the inverse of length $ 1/L $   and cyclotron frequency $ \omega $, taking different values of temperature $T$, in both APBC and PBC cases. Regions above the curves  are  the domains corresponding to the chirally broken region in parameter space, where any value of $g_s$ yields nonvanishing dynamical quark mass. The dependence of $g_s ^{(c)}$ with $\omega, T$ is similar to that reported in~\cite{Martinez:2018snm}: the thermal and magnetic effects compete  to suppress or enhance the chiral broken phase, respectively. As  the  magnetic  field strength  increases, $g_s ^{(c)} $ decreases and the  critical temperature  moves  toward  higher  values, manifesting the expected effect in literature known as magnetic catalysis. Looking especially at the periodicity of the boundaries, we see that for the APBC situation the critical curves depends on $1/L$ analogously to $T$: there are values of $T$ and $1/L$ where the critical coupling diverges, i.e. the  critical values of $T$ and $1/L$ above which there is no chiral  symmetry breaking. This is a consequence of  the parallelism between $1/L $ and $T$ within the APBC. In this sense, thermal and APBC size effects concur with magnetic ones. Passing on the PBC case, the outcome is unlike: $g_s ^{(c)} $ diminishes as $L$ decreases, causing the 
reinforcement of the chiral broken phase. Hence, the behavior of the critical coupling  presents a sizable dependence on the selection of the boundary conditions. In next Subsections we explore the consequences of this issue in more detail.

\subsection{Constituent quark masses and susceptibilities}


\begin{figure}
\centering
\includegraphics[width=1\columnwidth]{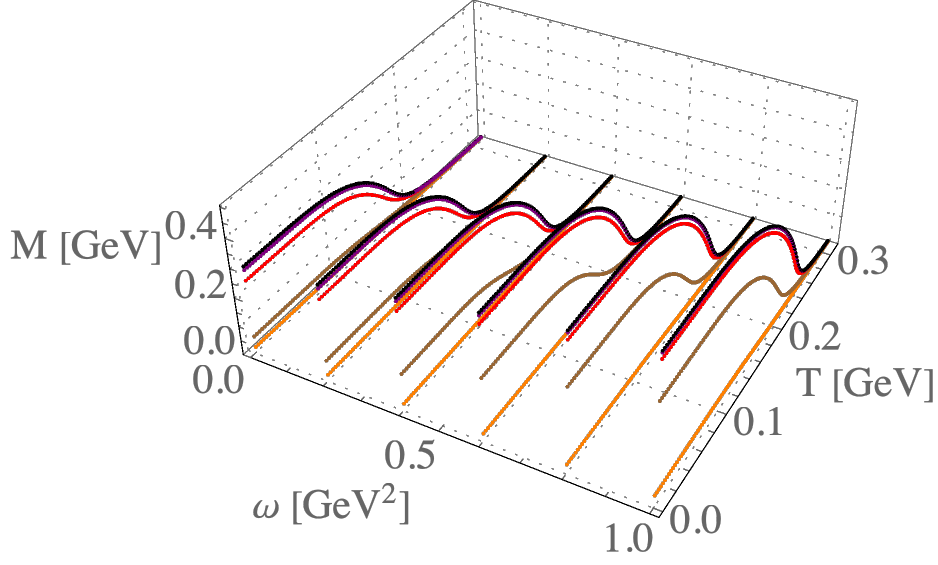} \\
\includegraphics[width=1\columnwidth]{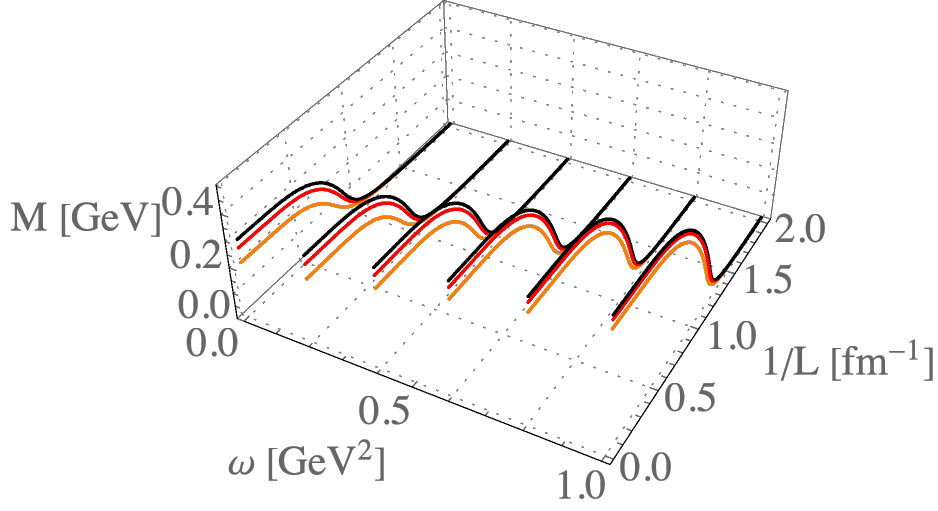}
\caption{Top panel: constituent quark mass as a function of temperature $ T $ and cyclotron frequency $ \omega $, taking different values of $L$ in APBC case. The black, purple, red, brown and orange points represent the results for $L= \infty, 2, 1.5, 1.0$ and $0.75$ fm, respectively. Bottom panel: constituent quark mass as a function of inverse of length $ 1/L $  in APBC case and cyclotron frequency $ \omega $, taking different values of temperature $T$). The black, red and orange points represent the results for $T = 0, 0.120$ and $0.150$ GeV, respectively.}
\label{QuarkMass1}
\end{figure}


\begin{figure}
\centering
\includegraphics[width=1\columnwidth]{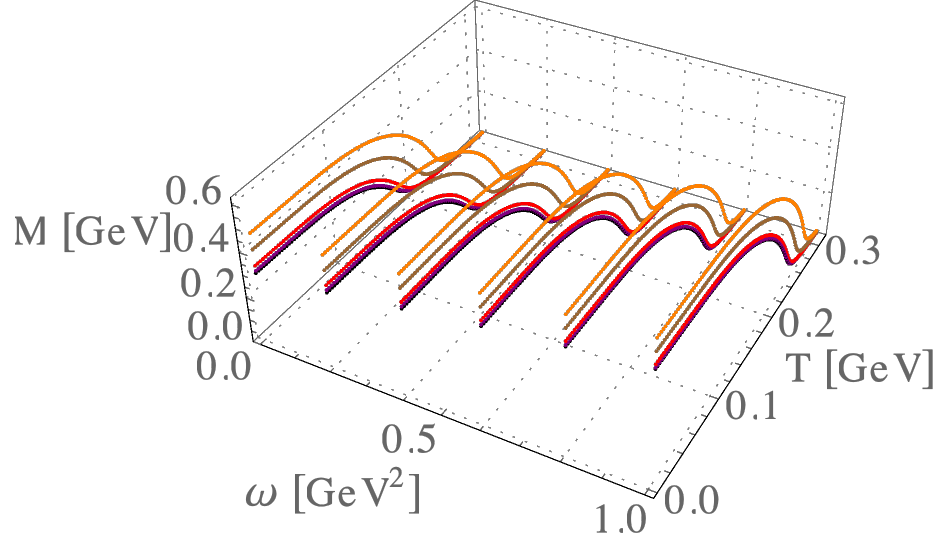}\\
\includegraphics[width=1\columnwidth]{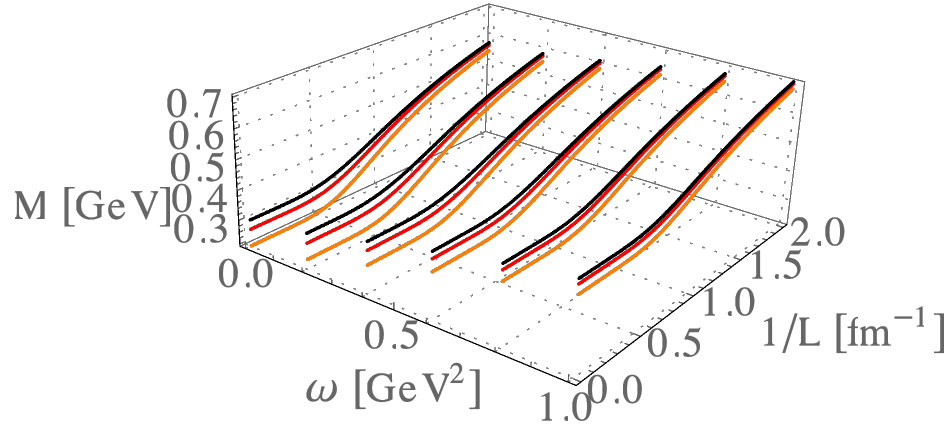}
\caption{The same as in Fig.~\ref{QuarkMass1}, but in PBC case.}
\label{QuarkMass1P}
\end{figure}


For the sake of completeness and comparison with other works, we start with the analysis of the constituent quark mass under the conjoint effects of boundaries, finite temperature and a magnetic background. With regard to other studies (see for example Refs.~\citep{Zhang:2016qrl,Wang:2017vtn,Mao:2018dqe,Avancini:2018svs,Zhao:2018nqa,Ghosh:2021dlo}), which  employed different treatments and methods to account for the finite volume or thermomagnetic effects, our aim is to discuss the combination/competition  of them in a unified way. 

In Figs.~\ref{QuarkMass1} and~\ref{QuarkMass1P} are plotted the values of $M$ that are solutions of the gap equation in Eq.~(\ref{massaH}) as a function of the the different variables, with spatial coordinates in APBC and PBC cases.  We see that at relatively smaller values of $T,  1/L$ and $\omega$ there is no appreciable modifications. In this region of thermodynamic variables, the vacuum mean-field approach appears as a good approximation. Notwithstanding, $M$ is hugely altered with the increase of any of these variables, and a strong dependence on the periodicity of boundary conditions appears. We start our analysis with the APBC situation and at sufficiently small magnetic field: it can be seen that at higher  $T$ (from approximately 100 MeV), $\mu$ (from approximately 100 MeV), $1/L$ (from approximately 0.3-0.5 fm${}^{-1}$), the constituent quark mass start to lower, and therefore the broken phase is inhibited and a crossover transition takes place. In particular, at certain values of $T$ and $1/L$ the dressed mass converges to the current quark mass. In this scenario, as expected the finite size and temperature cause similar effects to the chiral phase transition, due to the equivalent nature between $1/L $ and $T$, both using APBC.  

On the other hand, while in APBC case the presence of boundaries disfavors the maintenance of long-range correlations in a similar way to the finite temperature, in PBC situation the constituent quark masses acquire greater values with the decreasing of the size, causing a reverse effect compared to temperature. 
One can understand this difference in the phase structure more deeply by examining the behaviors of the $\theta$-functions in the chiral condensates shown in Eq.~(\ref{I1ETL}) and~(\ref{I1fTL})~\cite{Ishikawa:1996jb,Abreu:2020uxc}. 
The generalized Matsubara prescription (\ref{Matsubara}) states that the fermion fields with APBC cannot have a momentum less than $( {p}_{j}\rightarrow \bar{\omega} _{n_j}  \geq \pi / L_{j})$, with ${p}_{j}$ becoming larger for smaller values of $L_j$. Since the infrared contributions are relevant in the breaking of chiral symmetry, consequently in the chiral limit the quark condensate $\phi _f$ vanishes at a sufficiently small $L$  and hence the chiral symmetry is restored.  On the other hand, the PBC allow a zero value for ${p}_{j}$, which engenders no restoration of the symmetry as $L_j$ decreases. In fact, since the quark field $q_i (...,x_j,...) $ interacts with $q_i (...,x_j+L_j,...) $, in this context a finite $L_j$ yields a stronger interaction (caused by dimensional reduction), and the result of all correlations $\langle q_i(...,x_j,... ) q_i(...,x_j+L_j,...) \rangle$ is a higher value of $\phi _f$ with the decreasing of $L_j$.

It should be mentioned that is more frequent to find in literature of effective quark models APBC in spatial compactified directions for the quark fields~\cite{Abreu:2020uxc}. On the other hand, PBC are more often in lattice QCD simulations due to empirical minimization of finite-volume effects~\cite{Klein:2017shl} (see also Ref.~\cite{Magdy:2015eda}, which makes use of the $(2+1)-$flavor Polyakov linear sigma model with a purely mesonic potential).

Now we analyze the magnetic background effects. The augmentation of magnetic field strength stimulates the broken phase. This expected feature is comes from the magnetic catalysis effect. In the situation of APBC, $M$ acquires greater values and the increase of field strength induces even higher values of $T$ and $1/L$ for the crossover transition. Accordingly, the combination of finite size with APBC, finite temperature, finite chemical potential and magnetic effects generates a competition among them, since the last one yields its stimulation whereas the other ones suppress the broken phase.   
In contrast, when we consider the PBC, both drop of $L$ as well as increasing of $\omega$ cause greater values of $M$. 
Thereby, the conjunction of finite-size and magnetic effects on the phase structure has a hard dependence on the boundary conditions: while for APBC there is a concurrence between them, since the former inhibits the broken phase whereas the latter yields its enhancement; for PBC both effects cause stimulation of broken phase.

\begin{figure}
	\centering
	\includegraphics[width=1\columnwidth]{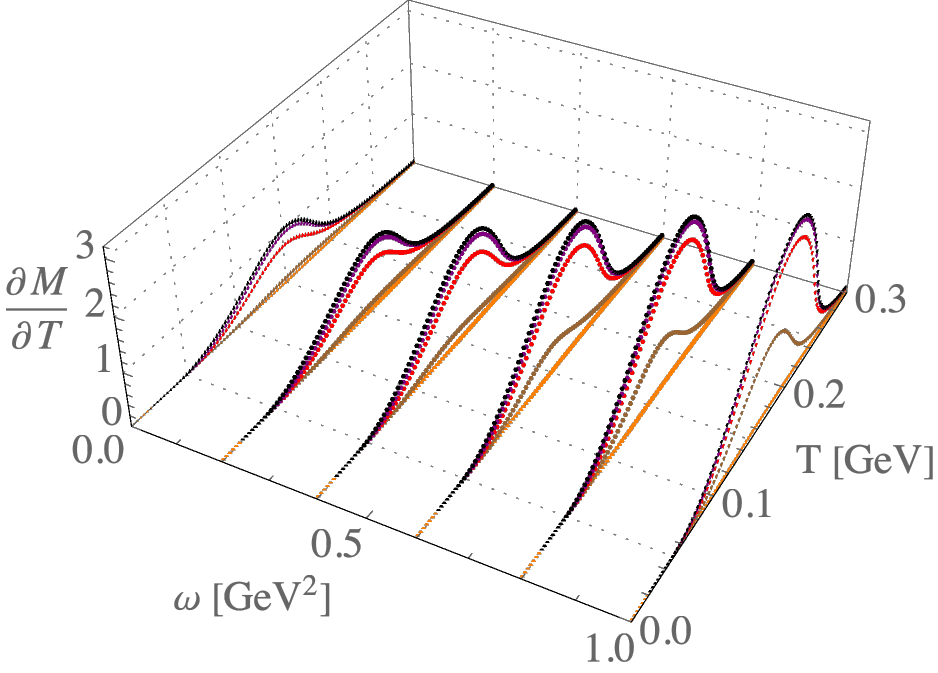}
	\\
	\includegraphics[width=1\columnwidth]{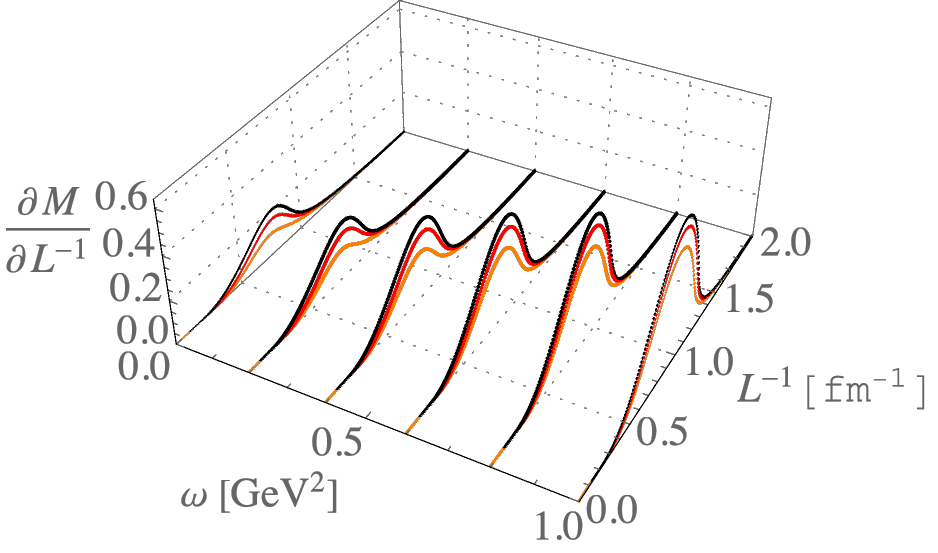}
	\caption{Top panel: chiral susceptibility of quark gas as a function of temperature $ T $ and cyclotron frequency $ \omega $, taking different values of $L$ in APBC case, at vanishing value of chemical potential $\mu$. The black, purple, red, brown and orange points represent the results for $L= \infty, 2, 1.5, 1.0$ and $0.75$ fm, respectively. Bottom panel: spatial susceptibility of quark gas as a function of $ 1/L $ and $ \omega $, taking different values of $T$, at $\mu=0$. The black, red and orange points represent the results for $T = 0, 0.120$ and $0.150$ GeV, respectively. }
\label{Susceptibilities}
\end{figure}

\begin{figure}
	\centering
	\includegraphics[width=1\columnwidth]{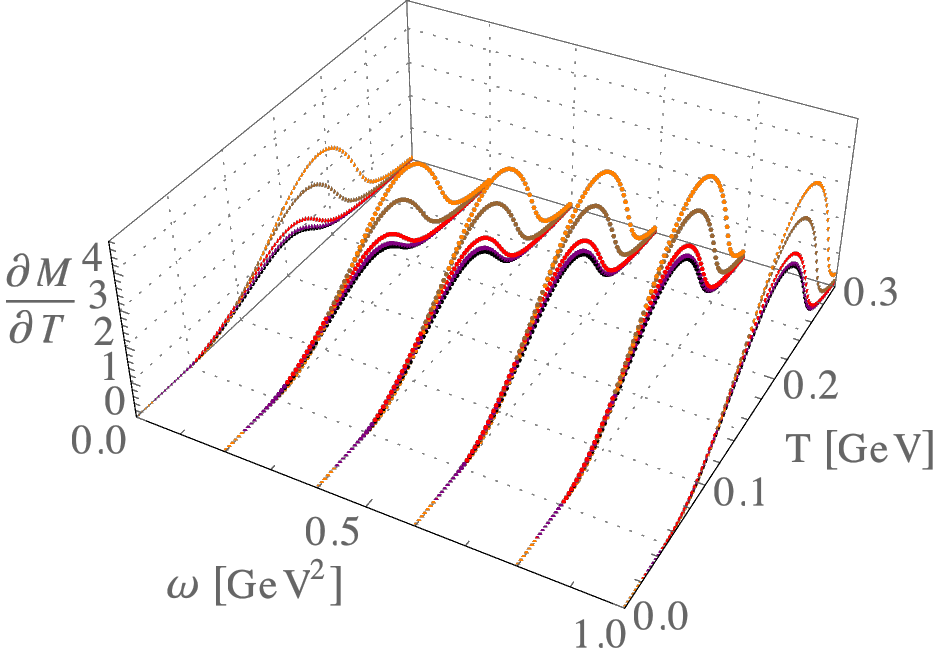}
	\\
	\includegraphics[width=1\columnwidth]{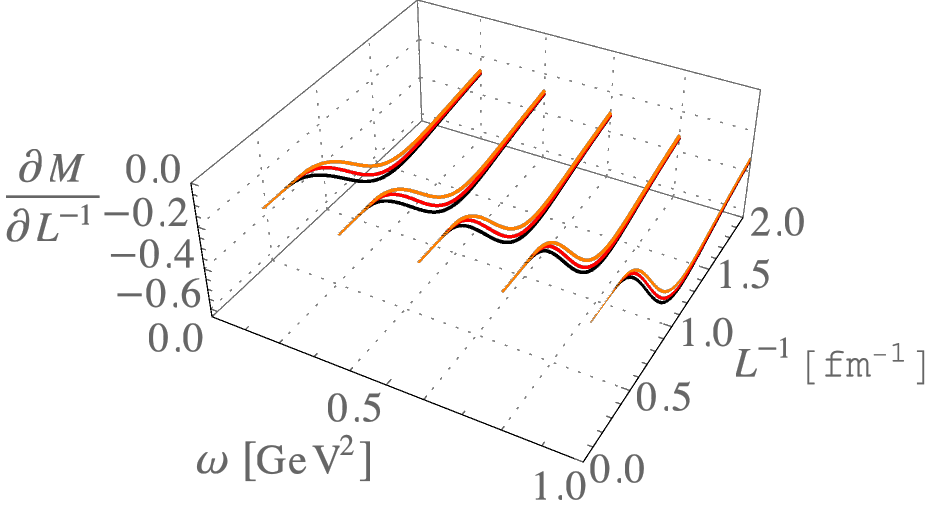}
	\caption{The same as in Fig.~\ref{Susceptibilities}, but in PBC case. }
\label{SusceptibilitiesPBC}
\end{figure}

To better characterize the phase structure, we also plot in Figs.~\ref{Susceptibilities} and~\ref{SusceptibilitiesPBC}  the chiral and spatial susceptibilities, defined as $\partial M / \partial T$ and $\partial M / \partial (L^{-1})$ respectively, taking different field strengths. Keeping in mind that for the APBC the peak indicates the point at which the chiral phase transition occurs, then the increase of $\omega$ augments the height of the peaks as well as the pseudo-critical temperature $T_c$ and pseudo-critical inverse of length $(L_c)^{-1}$. Besides, the decrease of $L$ yields the drop of the height of the peaks and $T_c$. In Table \ref{TableTcLc} we list some values of  $T_c$ and $(L_c)^{-1}$ as functions of the field strength. 
In the context of PBC, the peak in the chiral susceptibility moves to higher temperatures with increasing the magnetic field strength and drop of $L$. 
In other words, there is not a critical value of the size in which the symmetry is restored,  since the composite effect of magnetic background and boundary conditions in periodic case strengthens the broken phase.

\begin{center}
\begin{table}[ht]
\caption{Pseudo-critical temperature $T_c$ and inverse of pseudo-critical length $(L_c)^{-1}$
as functions of the field strength $\omega$, at different $L^{-1}$ in APBC and $T$. We adopt the notation as in Ref.~\cite{Zhang:2016qrl}: $T_c ^{\xi}$ and $(L_c ^{\xi})^{-1}$ denote the pseudo-critical values of the constituent quark mass $M$, obtained from the peaks of the chiral and spatial susceptibilities, respectively. All temperatures are in GeV, inverse of lengths in fm${}^{-1}$ and $\omega$ in GeV${}^{2}$. 
\label{TableTcLc}  } 
\begin{center}
\begin{tabular}{c|c|c||c|c|c}
\hline
\hline
$L^{-1}$ & $\omega$  & $ T_c ^{\xi}$ & $T$   & $\omega$ & $ (L_c ^{\xi})^{-1}$  \\ 
\hline
 \multirow{6}*{ 0 }     &  0.0    & 0.153 
 & \multirow{6}*{0}   &  0.0 & 0.78   \\
 							  &  0.2    & 0.156 & &  0.2 & 0.79  \\
 							  &  0.4    & 0.164 & &  0.4 & 0.84  \\
 							  &  0.6    & 0.175 & &  0.6 & 0.90  \\
 							  &  0.8    & 0.185 & &  0.8 & 0.95  \\
 							  &  1.0    & 0.195 & &  1.0 & 0.99  \\
\hline
 \multirow{6}*{0.67 }        &  0.0    & 0.150 
  & \multirow{6}*{120}   &  0.0 & 0.77   \\
 							  &  0.2    & 0.153 & &  0.2 & 0.78   \\
 							  &  0.4    & 0.162 & &  0.4 & 0.83  \\
 							  &  0.6    & 0.173 & &  0.6 & 0.89  \\
 							  &  0.8    & 0.183 & &  0.8 & 0.94  \\
 							  &  1.0    & 0.193 & &  1.0 & 0.98  \\
\hline
 \multirow{6}*{ $1.0$ }        &  0.0    & -  
   & \multirow{6}*{150}   &  0.0 & 0.74   \\
 							  &  0.2    & -     & &  0.2 & 0.75\\
 							  &  0.4    & 0.144 & &  0.4 & 0.82  \\
 							  &  0.6    & 0.150 & &  0.6 & 0.86 \\
 							  &  0.8    & 0.162 & &  0.8 & 0.92  \\
 							  &  1.0    & 0.175 & &  1.0 & 0.96  \\
\hline
\hline
\end{tabular}
\end{center}
\end{table}
\end{center}

\begin{figure}
\centering
\includegraphics[width=1\columnwidth]{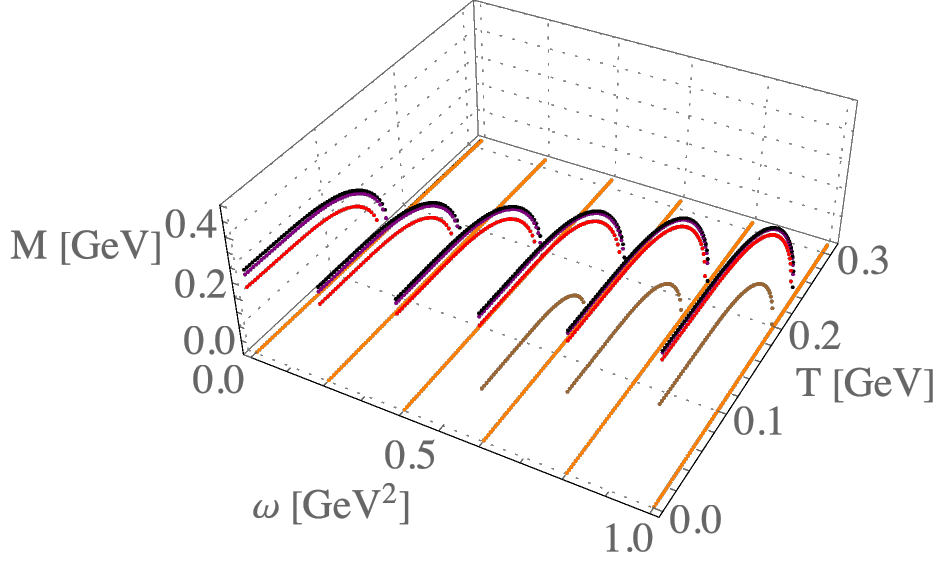} \\
\includegraphics[width=1\columnwidth]{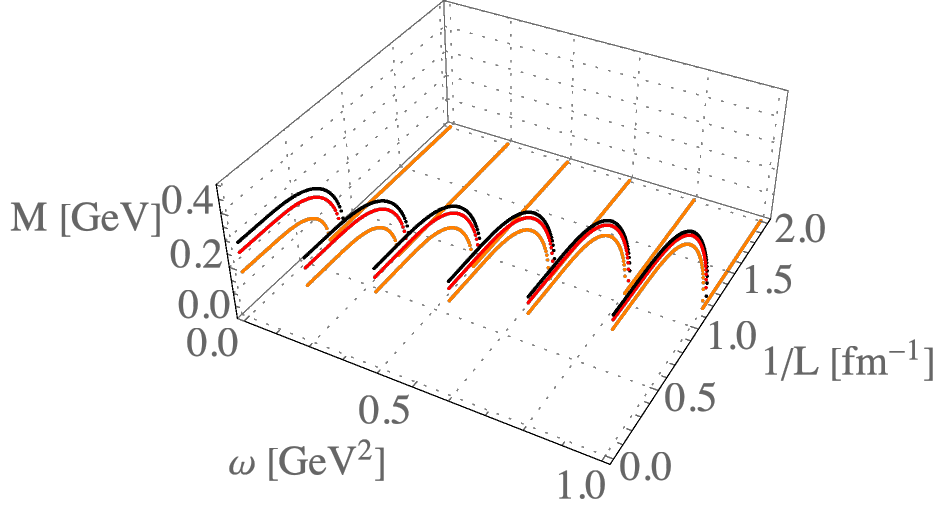}
\caption{Top panel: constituent quark mass $M$ in the chiral limit as a function of $ T $ and $ \omega $, taking different values of $L$ in APBC case, at $\mu =0$. The black, purple, red, brown and orange points represent the results for $L= \infty, 2, 1.5, 1.0$ and $0.75$ fm, respectively. Bottom panel: $M$ in the chiral limit as a function of $ 1/L $ and  $ \omega $, taking different values of $T$, at $\mu =0$. The black, red and orange points represent the results for $T = 0, 0.120$ and $0.150$ GeV, respectively.}
\label{QuarkMass3}
\end{figure}

\begin{figure}
\centering
\includegraphics[width=1\columnwidth]{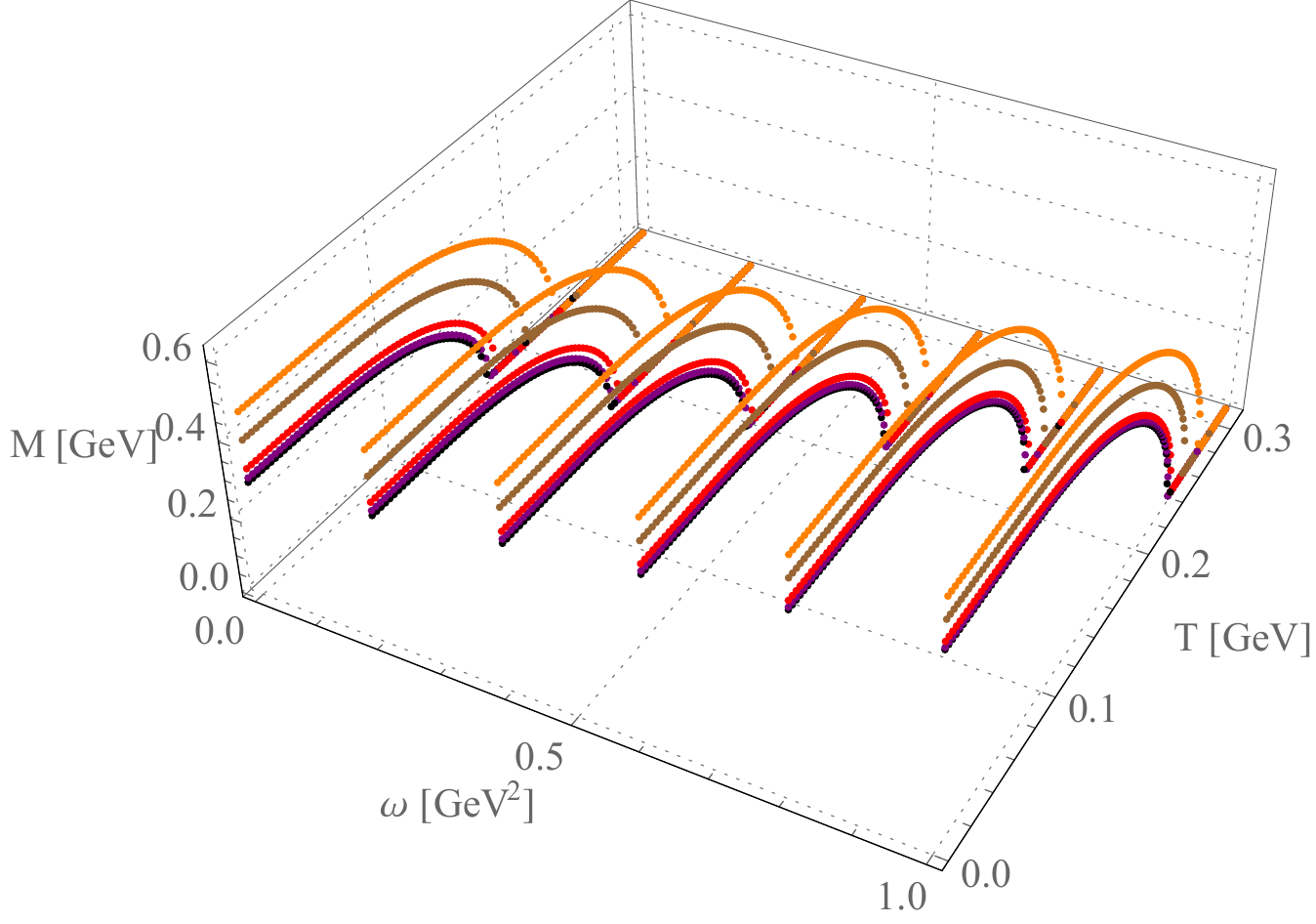} \\
\includegraphics[width=1\columnwidth]{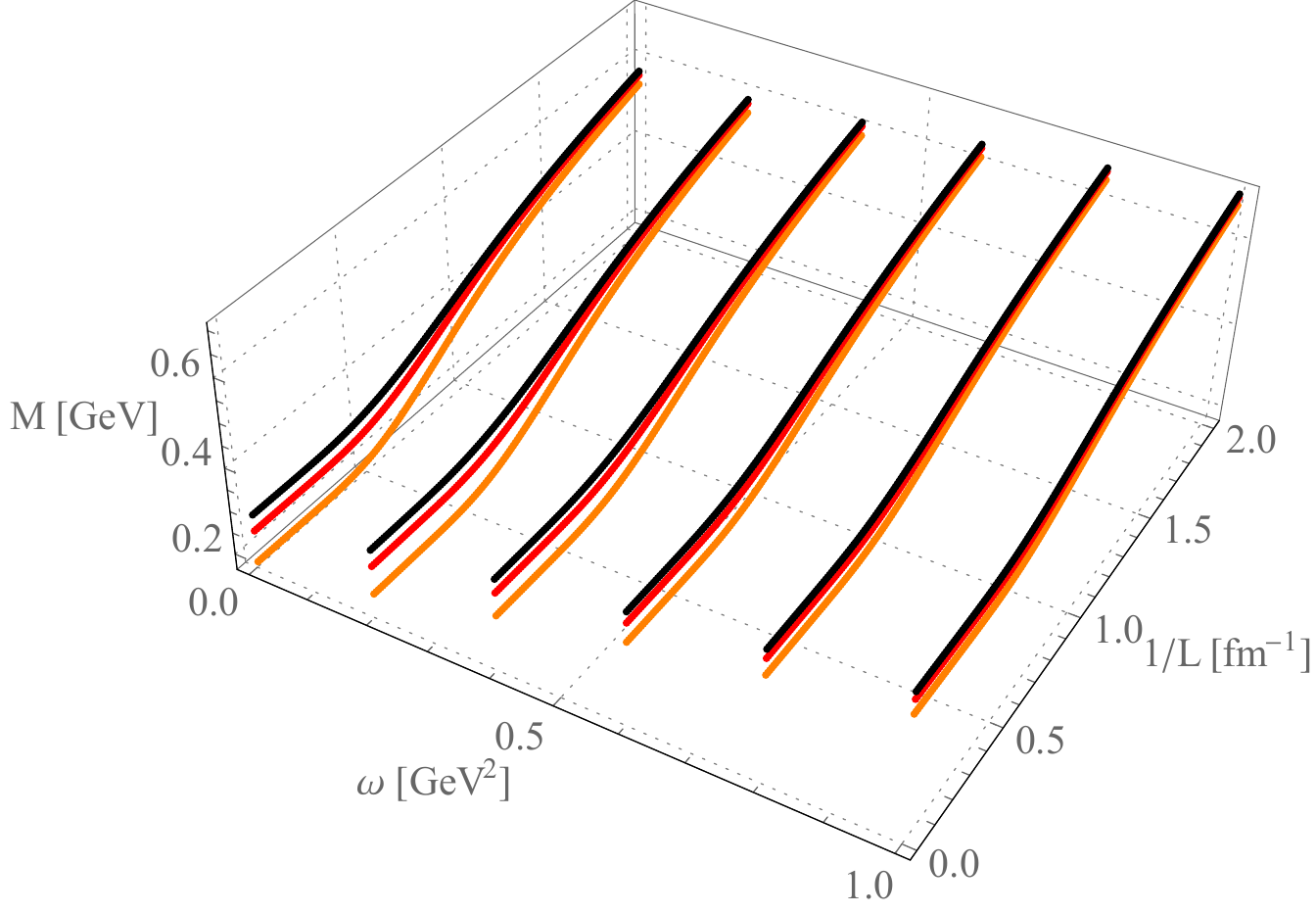}
\caption{The same as in Fig.~\ref{QuarkMass3}, but in PBC case.}
\label{QuarkMass3PBC}
\end{figure}

We finish this discussion on the phase structure with the chiral limit. This limit is achieved by fixing  $m=0$. Once this is an conceptual check of the chiral symmetry, without intention of practical purposes, we keep the values of the other parameters (effective coupling and ultraviolet cutoff) unchanged.  This choice gives the constituent quark mass $M = 0.285\,\mathrm{GeV} $ at $T, \mu, 1/L, \omega = 0$. The corresponding results are showed in Figs.~\ref{QuarkMass3} and~\ref{QuarkMass3PBC}. As it can be seen for the APBC case, in the chiral limit the dependence of the dressed quark mass with $1/L$  and $T$ shows a second order phase transition. In this situation the mentioned outcome becomes even more explicit: the increase of field strength forces the system to higher values of critical temperature and critical inverse of length. Interestingly, at sufficient $\omega$  the system with a certain $L$ suffers a transition from unbroken to broken chiral phase (see for example brown points in Fig.~\ref{QuarkMass3}). By contrast, for PBC higher values of both $\omega$ and  $1/L$  strengthen the broken phase.

\subsection{Meson Properties}

Now we can move on the properties of mesons propagating in a hot medium in the presence of boundaries and a magnetic background.

\begin{figure}
\centering
\includegraphics[width=1\columnwidth]{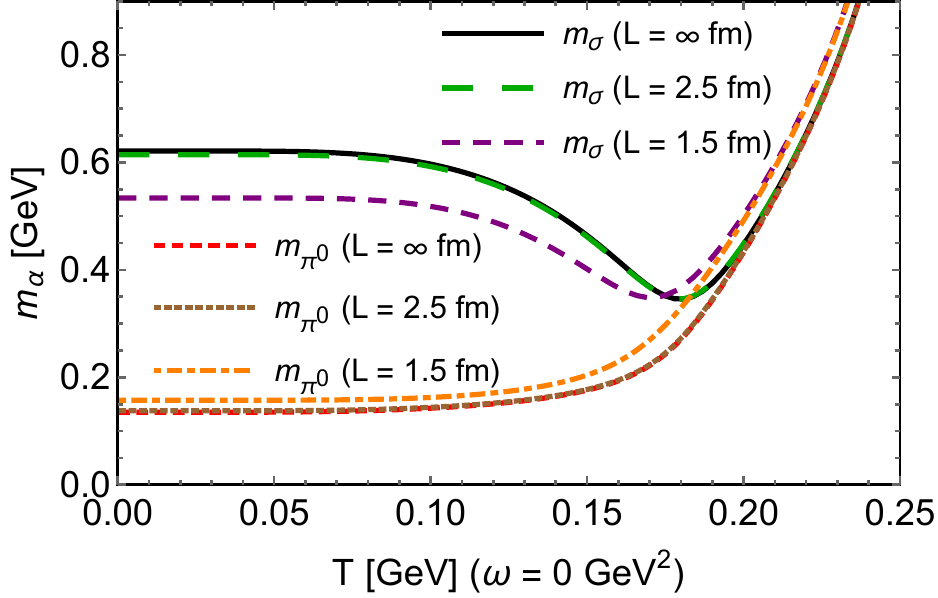}
\caption{$\pi ^0$ and $\sigma$ meson masses as functions of temperature in absence of magnetic field, taking different values of length $L$ in APBC case.}
\label{SigmaPionMasses1}
\end{figure}

\begin{figure}
\centering
\includegraphics[width=1\columnwidth]{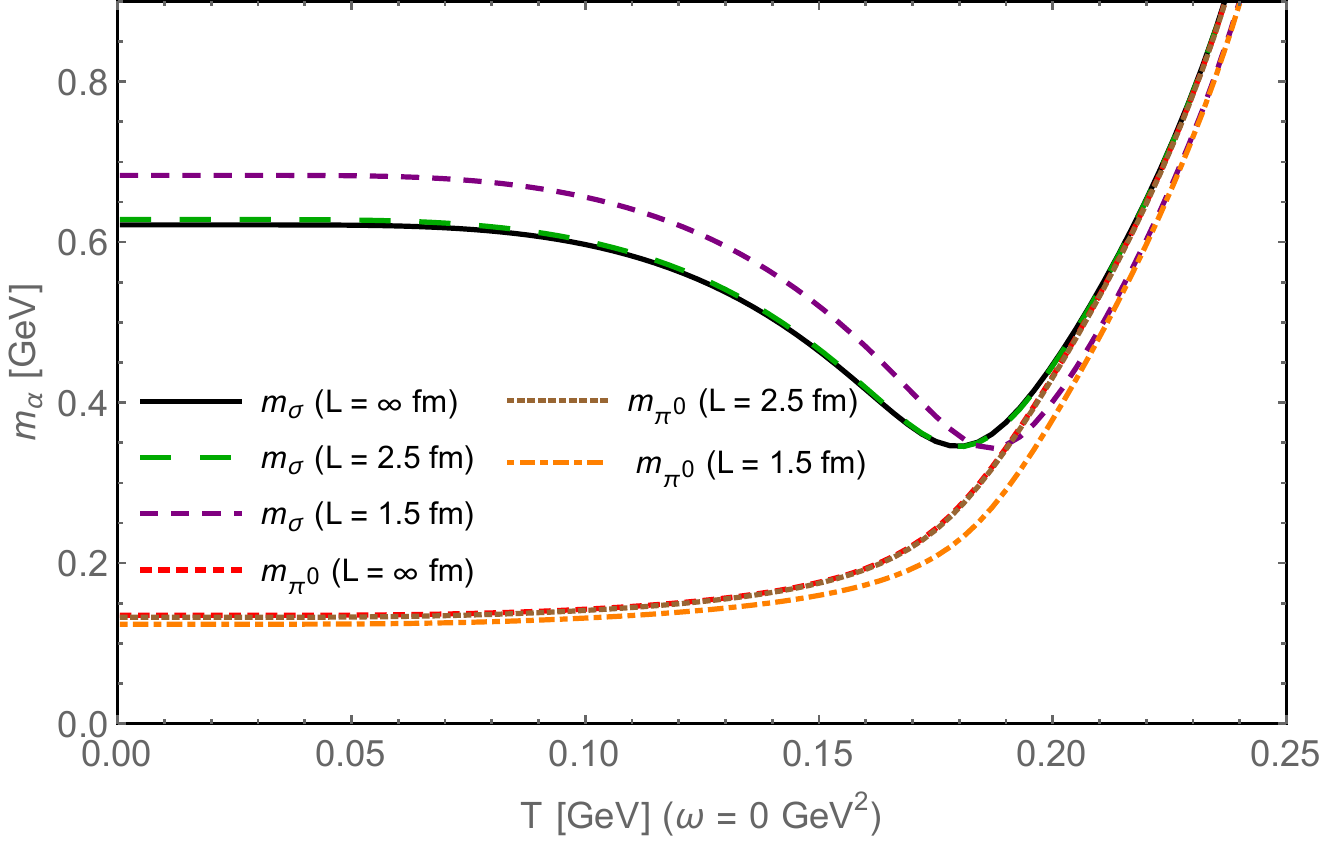}
\caption{
The same as in Fig.~\ref{SigmaPionMasses1}, but in PBC case. 
}
\label{SigmaPionMasses1PBC}
\end{figure}

First, we show in Figs.~\ref{SigmaPionMasses1} and~\ref{SigmaPionMasses1PBC} the behavior of $\pi ^0$ and $\sigma$ meson masses as functions of temperature in the case of absence of magnetic field, for different values of the length $L$, with spatial coordinates in APBC and PBC cases. In the bulk, the meson masses remain unchanged at smaller temperatures, but they are affected with the increase of $T$, suffering a crossover transition from the chiral symmetry broken phase to the chiral symmetry restored phase, and becoming degenerate. In particular, the $\sigma$ mass reaches a minimum at the critical point of chiral phase transition $ T_c ^{\xi}$, and then rises in the unbroken restored phase.  This is in qualitative agreement with previous findings, as those from~~\citep{Zhang:2016qrl,Wang:2017vtn,Mao:2018dqe,Avancini:2018svs,Zhao:2018nqa,Ghosh:2021dlo}. 
The finite size effects appear with a considerable dependence on the periodicity of boundary conditions. 
For the APBC, at sufficiently small values of $L$ the $\pi ^0$ and $\sigma$ masses are modified in a different way.  While the pion mass grows with the drop of the size, the mass of $\sigma$ decreases quite strong, in a phenomenon in which they become closer. This reflects the inhibition of chiral symmetry breaking  due to reduction of the size of the system. In this sense, the pseudo-critical temperature where the $\pi ^0$ and $\sigma$ meson masses start to degenerate also reduces with the decreasing of $L$. It is worthy remarking that an analogous effect is reported in Refs.~\cite{Zhao:2018nqa,Bhattacharyya:2012rp} by using distinct treatments.
Otherwise, in the context of PBC the $\pi ^0$ ($\sigma$) meson mass experiences an decrease (increase) as the size $L$ diminishes, and the pseudo-critical temperature augments with the drop of $L$.

\begin{figure}
\centering
\includegraphics[width=1\columnwidth]{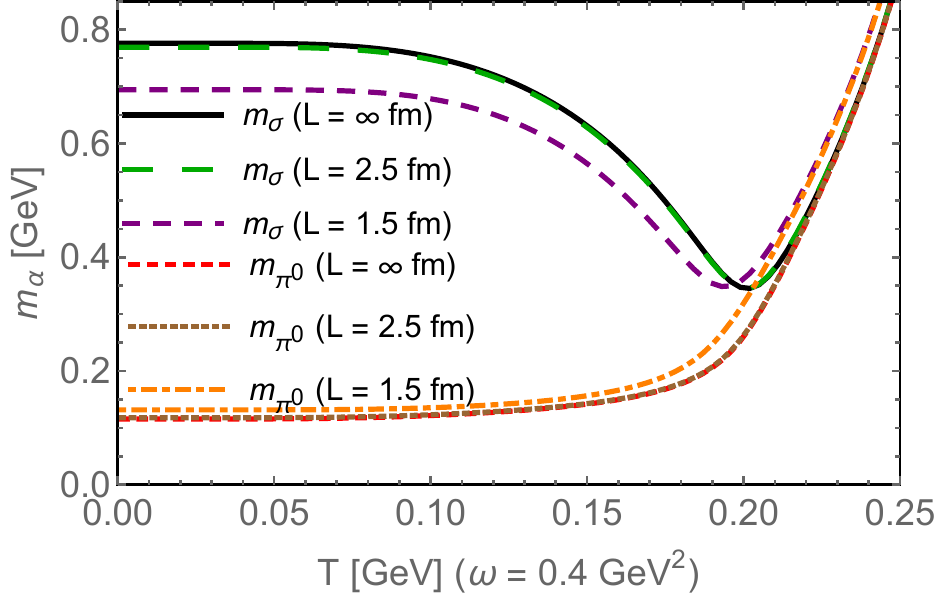}\\
\includegraphics[width=1\columnwidth]{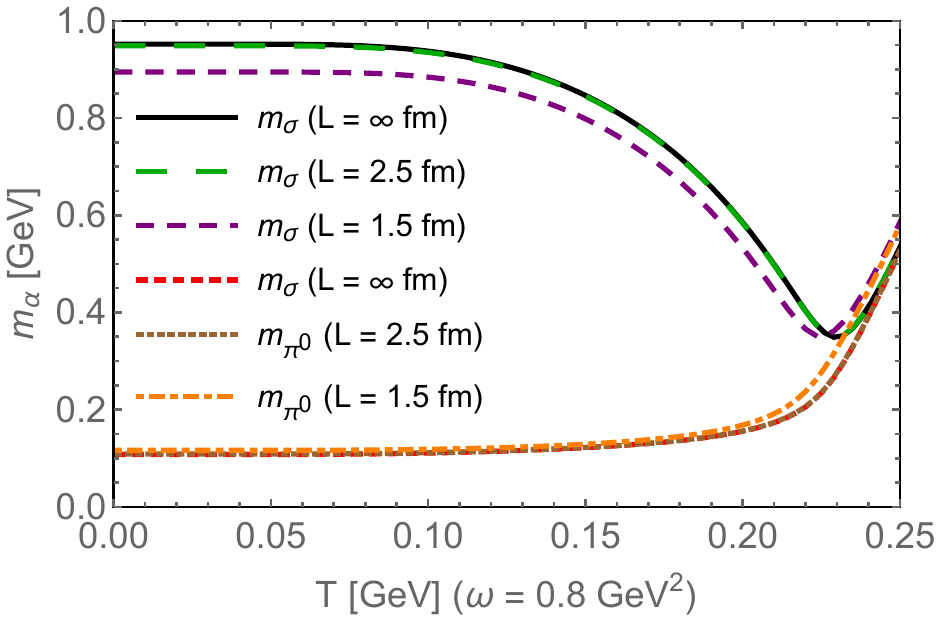}
\caption{$\sigma$ and $\pi ^0$ meson masses as functions of temperature, taking different values of length $L$  in APBC case and magnetic field strength $ \omega $. }
\label{SigmaPionMasses2}
\end{figure}

\begin{figure}
\centering
\includegraphics[width=1\columnwidth]{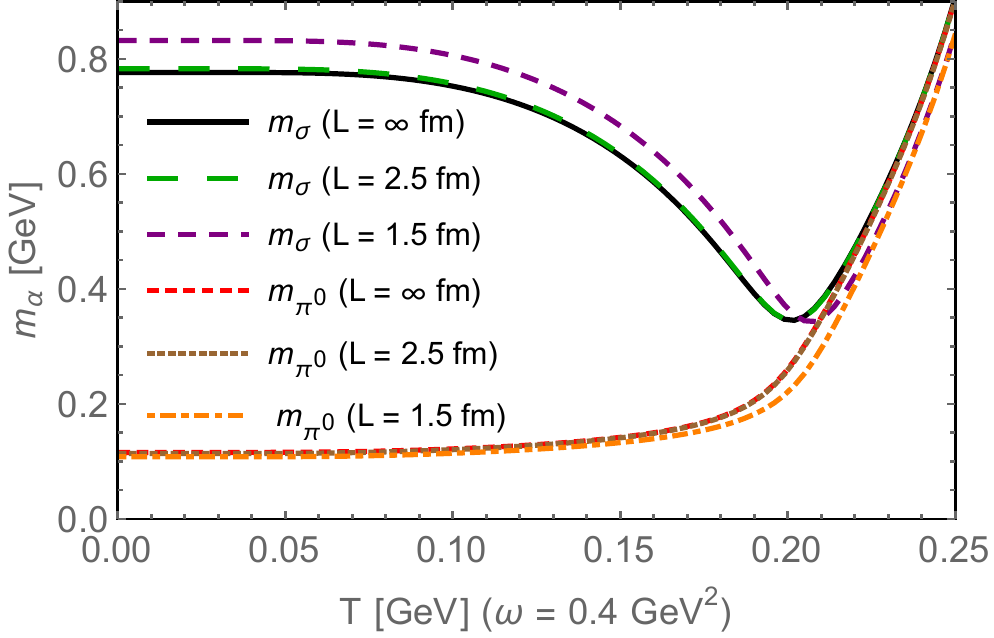}\\
\includegraphics[width=1\columnwidth]{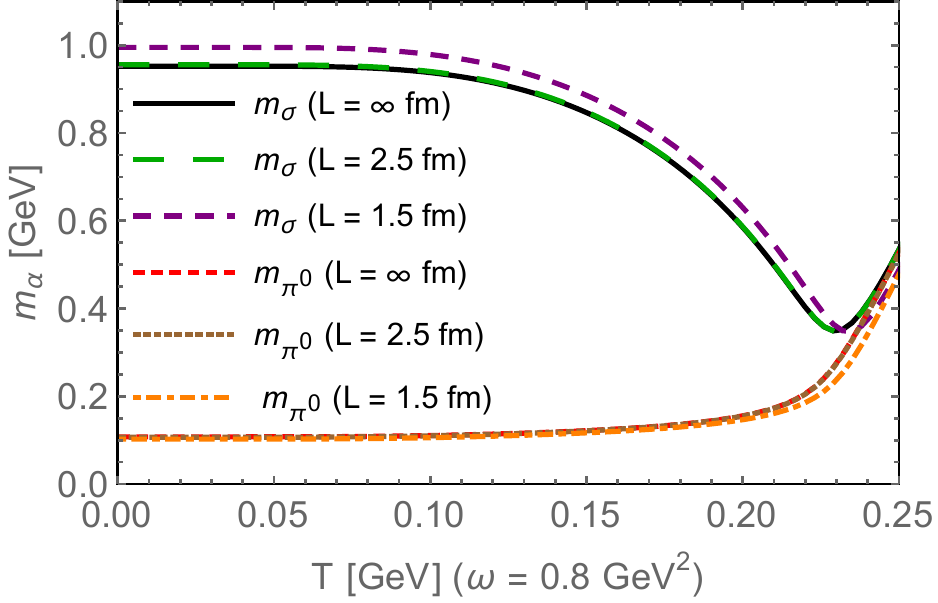}
\caption{The same as in Fig.~\ref{SigmaPionMasses2}, but in PBC case.}\label{SigmaPionMasses2PBC}
\end{figure}

\begin{figure}
\centering
\includegraphics[width=1\columnwidth]{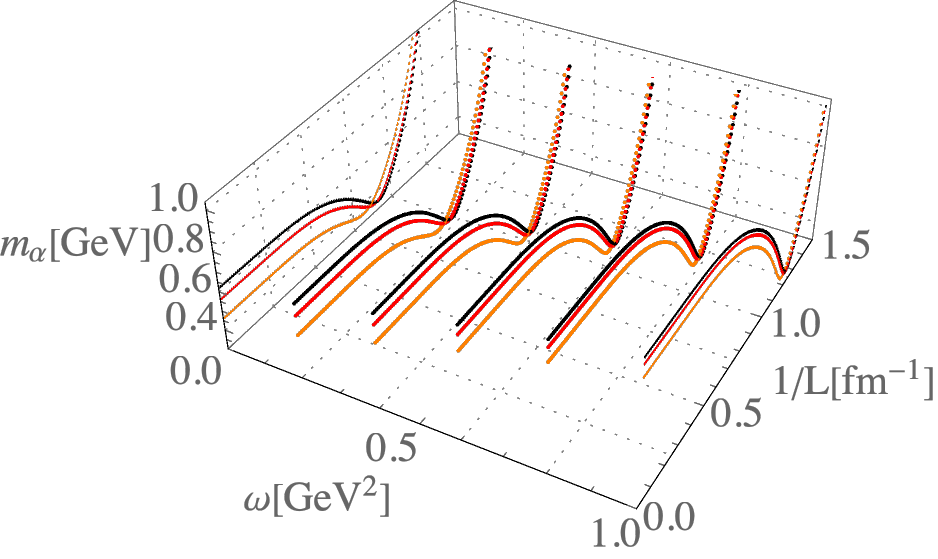}\\
\includegraphics[width=1\columnwidth]{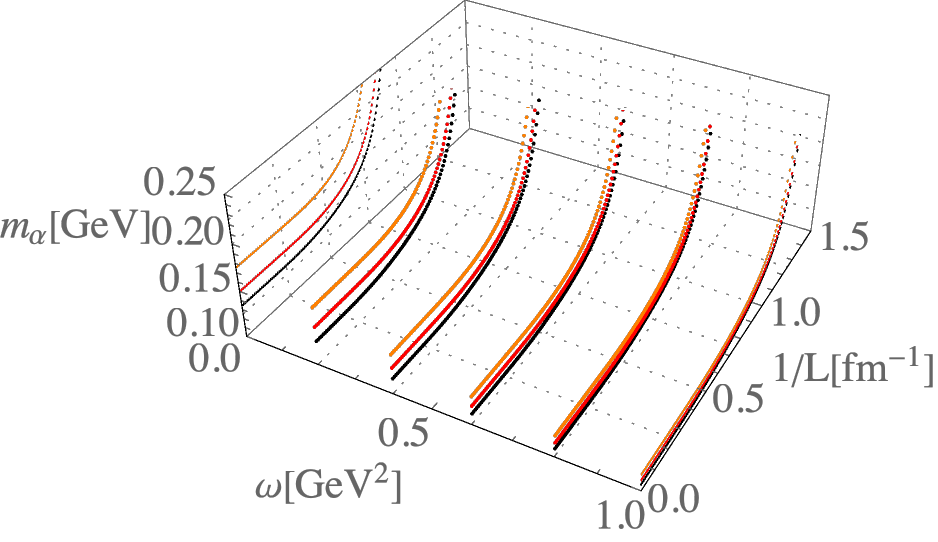}
\caption{$\sigma$ (top panel) and $\pi ^0$ (bottom panel) meson masses as functions of inverse of length $ 1/L $  in APBC case and cyclotron frequency $ \omega $, taking different values of temperature $T$. The black, red and orange points represent the results for $T = 0, 0.120$ and $0.150$ GeV, respectively. }
\label{SigmaPionMasses3}
\end{figure}

\begin{figure}
\centering
\includegraphics[width=1\columnwidth]{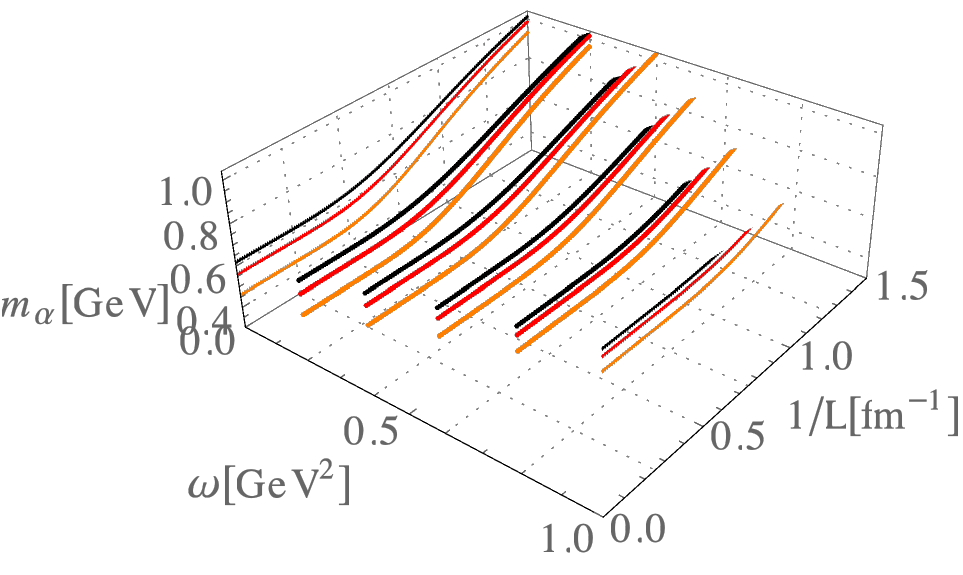}\\
\includegraphics[width=1\columnwidth]{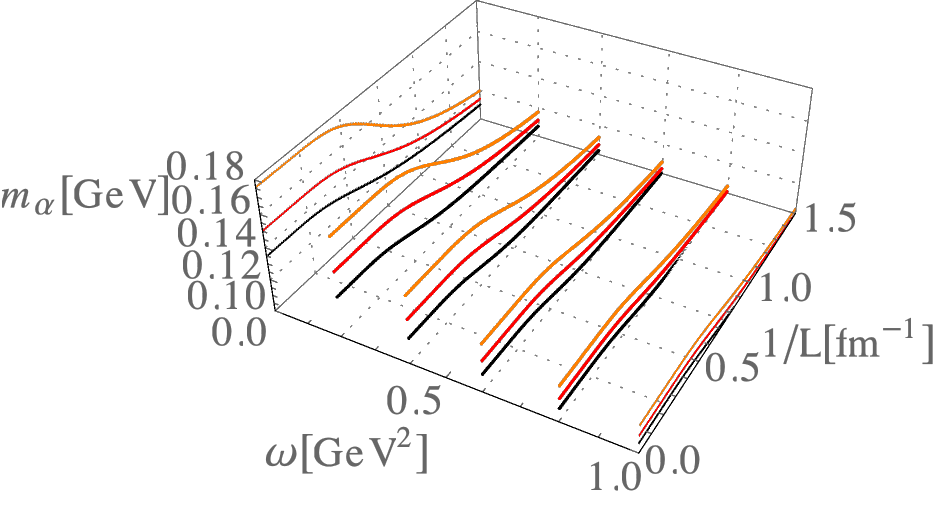}
\caption{The same as in Fig.~\ref{SigmaPionMasses3}, but in PBC case. }
\label{SigmaPionMasses3PBC}
\end{figure}

Now we focus on the evaluation of combined finite-size and thermomagnetic effects. In Figs.~\ref{SigmaPionMasses2}, ~\ref{SigmaPionMasses2PBC}, \ref{SigmaPionMasses3} and \ref{SigmaPionMasses3PBC} are plotted the calculated $\pi ^0$ and $\sigma$ meson masses a function of the relevant variables, including the cyclotron frequency $ \omega $. In the cases of spatial coordinates with APBC, the scalar meson $\sigma$ mass undergoes a competition between the effects of finite size and magnetic background: while the former lowers $m_{\sigma}$ and the pseudo-critical temperature at which $m_{\sigma}$ starts to degenerate,  the latter enhances them. Conversely, for the neutral pion the size effects enhance $m_{\pi ^0}$
but diminish the pseudo-critical temperature at which $m_{\pi ^0}$ begins to be degenerate, whereas the magnetic background reduces $m_{\pi ^0}$ and augments the pseudo-critical temperature. We notice that the dependence on the magnetic field and size of the neutral pion is weaker when compared to the one of the $\sigma$ meson. 
In the PBC scenario, the combination of the drop of $L$ and increasing of $\omega$ engender higher values for the scalar meson $\sigma$ mass and the pseudo-critical temperature; contrarily, the neutral pion mass diminishes.

\begin{figure}
\centering
\includegraphics[width=1\columnwidth]{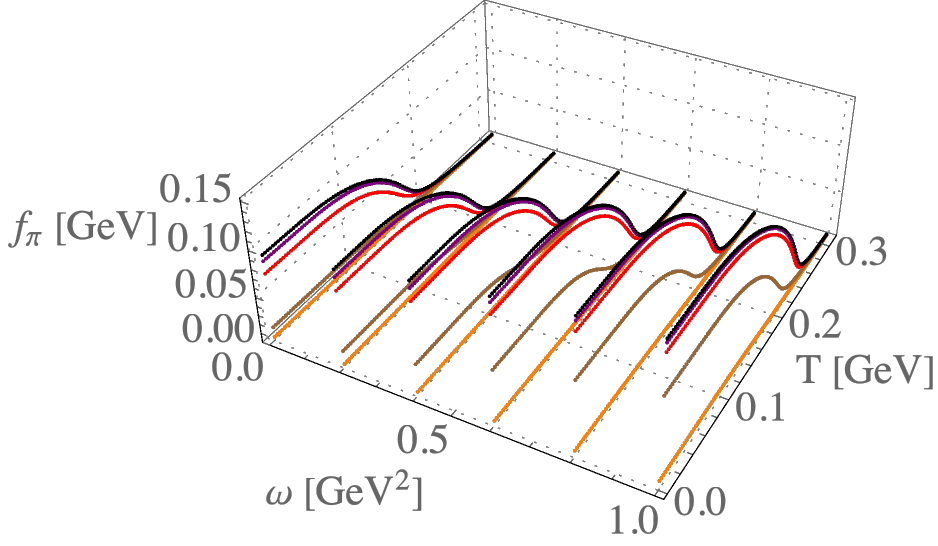}\\
\includegraphics[width=1\columnwidth]{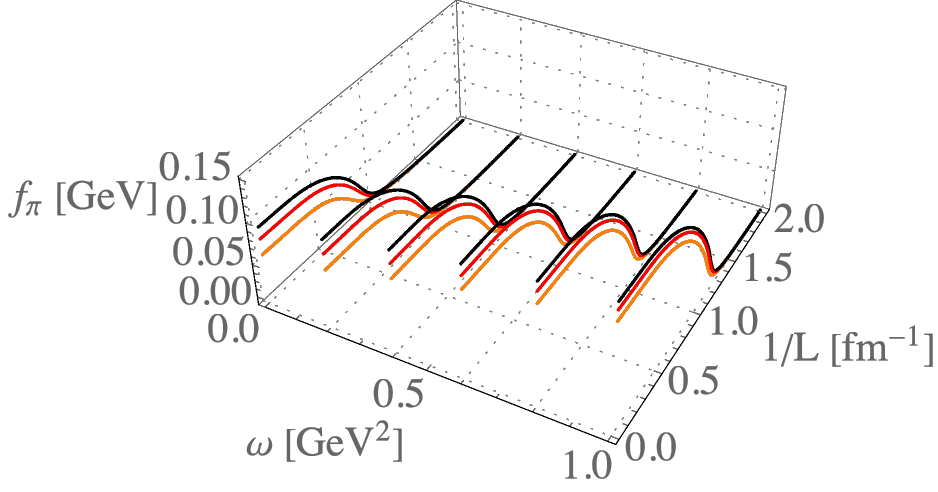}
\caption{Top panel: pion decay constant $f_{\pi}$ as a function of temperature $ T $ and cyclotron frequency $ \omega $, taking different values of $L$ in APBC case. The black, purple, red, brown and orange points represent the results in APBC case for $L= \infty, 2, 1.5, 1.0$ and $0.75$ fm, respectively. Bottom panel: $f_{\pi}$ as a function of $ 1/L $ and $ \omega $, taking different values of $T$, at $\mu=0$. The black, red and orange points represent the results for $T = 0, 0.120$ and $0.150$ GeV, respectively.}
\label{Piondecayconst}
\end{figure}

\begin{figure}
\centering
\includegraphics[width=1\columnwidth]{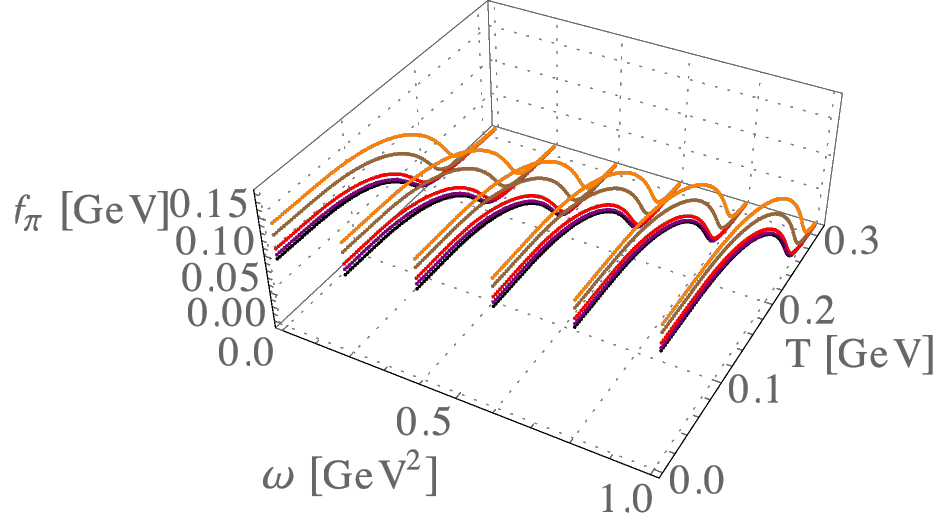}\\
\includegraphics[width=1\columnwidth]{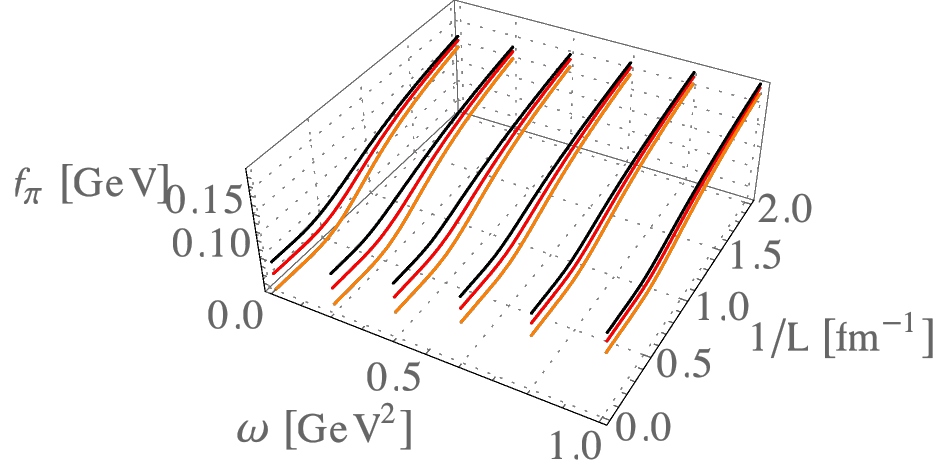}
\caption{The same as in Fig.~\ref{Piondecayconst}, but in PBC case. }
\label{PiondecayconstPBC}
\end{figure}

This analysis complemented with Figs.~\ref{Piondecayconst} and~\ref{PiondecayconstPBC}, where are plotted the values of the pion decay constant as functions of the relevant thermodynamic variables with spatial coordinates in PBC and APBC cases.  In the context of APBC, $f_{\pi}$ is large below the pseudo-critical temperature and inverse of the size and decreases with increasing $T$ and $1/L$. This outcome is in accordance with that reported in Ref.~\cite{Bhattacharyya:2012rp}, and it is an symptom of the restoration of the chiral symmetry, since  $f_{\pi}$ is proportional to the divergence of the chiral current $J_5 ^{a \mu} = \bar{q} \gamma ^{\mu} \gamma _5 \frac{\tau ^a}{2} q$.  Conversely, $f_{\pi}$ acquires higher values with the growth of the field strength, due to the magnetic catalysis, manifesting its concurrence with the finite size (APBC) and temperature effects. 
By contrast, the case of PBC makes the pion decay constant reach higher values with both increasing of $1/L$ and $\omega$. 
As the other quantities studied above, the behavior of $f_{\pi}$  has a hard dependence on the periodicity of the boundary conditions.

\begin{figure}
\centering
\includegraphics[width=1\columnwidth]{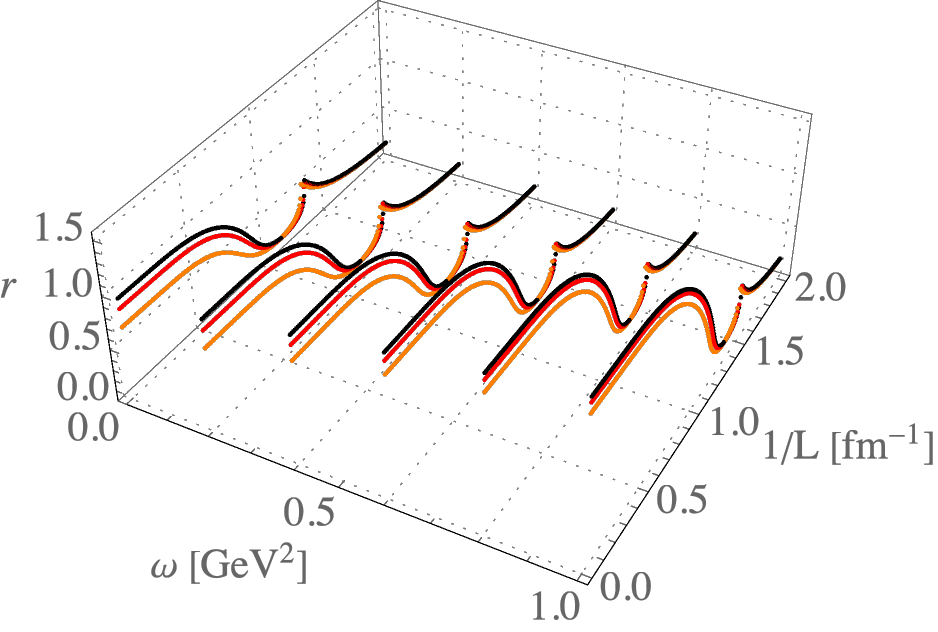}
\caption{Ratio $r$ defined in Eq.~(\ref{Ratio}) as a function of $ 1/L $ in APBC and $ \omega $, taking different values of $T$. The black, red and orange points represent the results for $T = 0, 0.120$ and $0.150$ GeV, respectively.}
\label{FigRatio}
\end{figure}

\begin{figure}
\centering
\includegraphics[width=1\columnwidth]{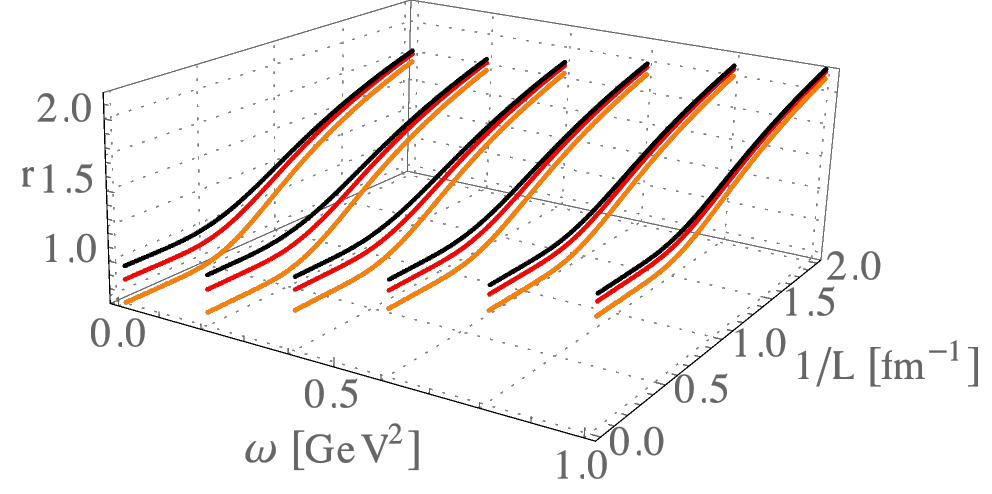}
\caption{The same as in Fig.~\ref{Piondecayconst}, but in PBC case.}
\label{FigRatioPBC}
\end{figure}

It is interesting to check the Gell-Mann–-Oakes–-Renner (GMOR) relation, which relates the  mass and decay constant of the pion to the current quark mass and quark condensate, and tells us how the dynamical quantities must evolve in such a way to satisfy it (see Ref.~\cite{Zhang:2016qrl,Fu:2009zs} for a detailed discussion). In the lowest order of chiral expansion, it is given by 
\begin{eqnarray}
f_{\pi} ^2 m_{\pi} ^2 = - 2 m \langle \bar{q} q\rangle . 
\label{GOR1} 
\end{eqnarray}
The two flavors are taken into account in the quark condensate. The influence of the thermodynamic variables $T, 1/L$ and $\omega$ in this relation can be studied through the ratio~\cite{Zhang:2016qrl,Fu:2009zs}
\begin{eqnarray}
r  = \frac{f_{\pi} ^2 m_{\pi} ^2}{- 2 m \langle \bar{q} q\rangle} . 
\label{Ratio} 
\end{eqnarray}
At $T, 1/L, \omega \rightarrow 0$, the GMOR relation is very well preserved $(r \approx 1)$, which is a demonstration of the dynamical chiral symmetry breaking. 
However, if for example temperature increases the GMOR relation should break down $(r \neq 1)$, as the chiral symmetry is restored. 
To verify how the ratio $r$ is affected by the finite size and magnetic effects, in Figs.~\ref{FigRatio} and~\ref{FigRatioPBC} are plotted the ration $r$ as a function of $ 1/L $ and $ \omega $, taking different values of temperature and with spatial coordinates in PBC and APBC cases. As expected, the increase of $T$ up to sufficiently higher values enforces the breaking down of the GMOR relation due to the start of the restoration of the chiral symmetry. But at moderate values of $T$, $\omega$ and $1/L$, the ratio does not deviate from 1.  More interestingly, in the situation of APBC, there is a range of values of $1/L$ at which $r$ reaches local minimum and maximum. This behavior is similar to the one discussed for finite temperature effects in~\cite{Zhang:2016qrl}. So, at enough smaller sizes the GMOR relation is violated, and this fluctuation can be interpreted as a sign of the chiral phase transition. In other words, the presence of antiperiodic boundaries induces the restoration of dynamical chiral symmetry, analogously to the thermal effect. This fact reflects the  presupposition of the physical equivalence among antiperiodic compactified spatial and imaginary time coordinates. We also notice that sufficiently high field strength affects the region near the critical region of $T$ and $1/L$, yielding higher $T$ and $1/L$ at which $r$ reaches local minimum and  maximum.
In a different way, the context of PBC does not engenders a region in $r$ with a local minimum and a maximum; it suffers an augmentation as $1/L$ increases.

%
\section{Concluding remarks}

In this work we have examined the competition between the finite volume and magnetic effects on the properties of neutral mesons in a hot medium, in the context of the Nambu--Jona-Lasinio model. By using the mean-field approximation and the Schwinger proper time method in a toroidal topology with antiperiodic conditions, we have investigated the gap equation solutions and meson observables like the $\pi ^0$ and $\sigma$ meson masses and pion decay constant under the change of the size, temperature and strength of external magnetic field. Our findings have shown that these observables are strongly affected by the combined effects of relevant variables, depending on their range,  and also by the periodicity of the boundary conditions. 

We have seen that the conjunction of boundaries with antiperiodic conditions and magnetic effects generates a competition among them, since the last one yields  an enhancement of the dynamical breaking of chiral symmetry (higher values of the constituent quark mass $M$ and pseudo-critical temperature $ T_c ^{\xi}$), whereas the former one favors the restoration of the chiral symmetry (smaller $M$ and $ T_c ^{\xi}$). In contrast,  while for APBC there is a concurrence between finite-size and magnetic effects, for PBC both effects stimulate the broken phase.

Our evaluation of the conjoint finite-size and thermomagnetic effects on the meson properties has revealed two distinct patterns between the $\pi ^0$ and $\sigma$ meson masses. In the APBC, whereas $m_{\sigma}$ and its pseudo-critical temperature decreases with the presence of boundaries and increases with a magnetic background, $m_{\pi ^0}$ is increased at smaller $L$ and diminishes with the field strength. In other words, we have found that in an environment with sufficiently restricted volume (and similarly with high temperatures) and high magnetic field, the masses of these chiral partners experience simultaneously the tendency to be nearer each other due to the boundaries as well as farther because of the magnetic background, and the net result will depend on the balance of these competing effects.
We have remarked the neutral pion presents a feeble dependence on these variables when compared to the $\sigma$ meson. 
In the PBC scenario, the combination of the descrease of $L$ and increasing of $\omega$ engender higher values for the scalar meson $\sigma$ mass and the pseudo-critical temperature; contrarily, the neutral pion mass diminishes. 
Concerning the pion decay constant, $f_{\pi}$ acquires higher values with the growth of the field strength, due to the magnetic catalysis, manifesting its concurrence with the APBC finite size and temperature effects. Conversely, the case of PBC makes $f_{\pi}$ reach higher values with both increasing of $1/L$ and $\omega$.

We should notice that the dependence of the findings obtained above on the regularization procedure, parametrization choice, and finite-size prescription can not be underrated. Obviously, a different set of parameters considered as input will alter the magnitudes of the constituent quark mass,  meson properties and ranges of $(T,L, \omega)$ which engender changes on their behavior. Thus, the comparison with existing literature must be done carefully, keeping in mind the techniques employed. 

Finally, we mention that the findings outlined above can provide insights on the finite-volume and magnetic effects that are relevant in the scenario of quark matter produced in experiments as heavy-ion collisions or in lattice simulations. Subsequent studies are essential to explore the efficacy of the proposed framework and estimate the range of size of the system at which the bulk is a reasonable approximation. Likewise, the inverse magnetic catalysis will deserve our attention in a future work.

\acknowledgments

This work was partially supported by the following grants (L.M.A.): the Brazilian CNPq (contracts 309950/2020-1 and 400546/2016-7) and FAPESB (contract INT0007/2016).  

%

\end{document}